\documentclass{article} 
\usepackage{iclr2024_conference,times}

\usepackage{amsmath,amsfonts,bm}

\def\eqref#1{equation~\ref{#1}}

\def\1{\bm{1}}

\DeclareMathAlphabet{\mathsfit}{\encodingdefault}{\sfdefault}{m}{sl}
\SetMathAlphabet{\mathsfit}{bold}{\encodingdefault}{\sfdefault}{bx}{n}

\usepackage{hyperref}
\usepackage{url}
\usepackage{graphicx}
\usepackage{booktabs}
\usepackage{subcaption}
\usepackage{wrapfig}
\usepackage{colortbl} 
\usepackage{xcolor} 
\usepackage{multirow} 
\usepackage{pifont}

\newcommand{\cmark}{\ding{51}}%
\newcommand{\xmark}{\ding{55}}%

\title{Integrating the Expected Future in Load Forecasts with Contextually Enhanced Transformer Models}

\author{Raffael Theiler,  Leandro Von Krannichfeldt \\
Intelligent Maintenance and Operations Systems (IMOS)\\
École Polytechnique Fédérale de Lausanne (EPFL)\\
Lausanne, CH-1015, Lausanne \\
\texttt{\{raffael.theiler, leandro.vonkrannichfeldt\}@epfl.ch} \\
\And
Giovanni Sansavini \\
Department of Mechanical and Process Engineering \\
Eidgenössische Technische Hochschule (ETH) \\
Zürich, CH-8092, Zürich \\
\texttt{sansavig@ethz.ch} \\
\AND
Michael F. Howland \\
Department of Civil and Environmental Engineering \\
Massachusetts Institute of Technology (MIT) \\
77 Massachusetts Avenue, Cambridge, MA 02139 \\
\texttt{mhowland@mit.edu} \\
\And
Olga Fink \\
Intelligent Maintenance and Operations Systems (IMOS)\\
École Polytechnique Fédérale de Lausanne (EPFL)\\
Lausanne, CH-1015, Vaud \\
\texttt{olga.fink@epfl.ch} \\
}

\iclrfinalcopy %
\begin{document}

\maketitle

\begin{abstract}

Accurate and reliable energy forecasting is essential for power grid operators
who strive  to minimize  extreme forecasting errors 
that pose significant operational challenges and incur high intra-day trading costs.
Incorporating planning information -- such as anticipated   
user behavior, scheduled events or timetables --
provides substantial contextual information to enhance forecast  accuracy and reduce the occurrence of large forecasting errors. 
Existing approaches, however, lack the flexibility 
to effectively integrate  both dynamic, forward-looking contextual inputs and historical data.
In this work, 
we conceptualize forecasting as a combined forecasting-regression task, formulated  
as a sequence-to-sequence prediction problem, and introduce contextually-enhanced transformer models designed to leverage all contextual information effectively.  
We demonstrate the effectiveness of our approach through a primary case study on nationwide railway energy consumption forecasting, where integrating contextual information into transformer models, particularly timetable data, resulted in a significant  average mean absolute error  reduction of    26.6\%. 
An auxiliary case study on building energy forecasting, leveraging planned office occupancy data, further illustrates the generalizability of our method, showing an average reduction
of 56.3\% in mean absolute error. Compared to other state-of-the-art methods, our approach consistently outperforms existing models, underscoring the value of context-aware deep learning techniques in energy forecasting applications.

\end{abstract}

\vspace{1em}


\section{Introduction}

Accurate energy forecasting is crucial for the effective and reliable operation of power grids
\cite{mahzarniaReviewMeasuresEnhance2020}. 
The precision of these forecasts impacts various aspects, including
long-term planning,
grid stability, 
operational efficiency, 
economic performance
and environmental sustainability
\cite{hongEnergyForecastingReview2020, 
foseEmpoweringDistributionSystem2024}.

For grid operators, precise forecasts ensure the critical balance between electricity supply and demand, thereby preventing blackouts and minimizing the need for costly emergency interventions  
\citep{foseEmpoweringDistributionSystem2024, 
ahmadReviewRenewableEnergy2020, 
anandBottomupForecastingApplications2023}.
Economically, accurate predictions help reduce operational costs by optimizing energy procurement and decreasing
reliance on expensive intra-day trading
\citep{sweeneyFutureForecastingRenewable2020}. 
From an environmental perspective, reliable forecasts facilitate the integration of renewable energy sources, 
promoting cleaner and more sustainable energy systems,
while also reducing the dependence on resource-intensive storage technologies
\citep{ahmadReviewRenewableEnergy2020}.
However, significant inaccuracies in predictions 
-- particularly those caused by large outliers -- can undermine these benefits and amplify challenges across all aspects of energy management.

The ongoing liberalization and decentralization of the energy sector, 
has significantly transformed the landscape of energy forecasting  
\citep{foseEmpoweringDistributionSystem2024,
ahlqvistSurveyComparingCentralized2022}.
Traditionally, energy generation and consumption were centralized, involving a limited number of large-scale producers and aggregated consumers. 
Forecasting at this centralized level \citep{wijayaClusterbasedAggregateForecasting2015, plaumAggregatedDemandsideEnergy2022} is relatively straightforward,
with simple forecasting models performing  comparably  to sophisticated ones
\citep{kusterElectricalLoadForecasting2017}. 
This effectiveness was largely due to the averaging effect of diverse local and regional factors
\cite{alemazkoorSmartMeterBigData2022,
ninagawaPredictionAggregatedPower2016,  anandBottomupForecastingApplications2023}.

The recent shift towards decentralization has significantly  increased  the number of 
stakeholders requiring precise energy forecasts
\cite{disilvestreHowDecarbonizationDigitalization2018}.
The rapid growth of distributed energy resources (DERs)  \cite{chenModernizingDistributionSystem2017}, is driven by advancements in renewable energy technologies, government policies for heating and mobility electrification, and the increasing demand for sustainable and resilient energy systems \cite{ExecutiveSummaryUnlocking}. For instance, by 2027, the US DER market is projected to double, adding 262 gigawatts (GW) of new DER and demand flexibility capacity \cite{plaumAggregatedDemandsideEnergy2022, mackenzieUSDistributedEnergy2023}, while the distributed storage market is expected to more than quadruple \cite{hertz-shargelTransformationUSDistributed2023}. Additionally, the global microgrid market was projected  to grow  fivefold between 2019 and 2028 \cite{lenhartMicrogridDecisionmakingPublic2021} and reached an installed capacity of 10 GW by 2022 \cite{mackenzieUSMicrogridMarket2023}. The primary driver for this trend is the increasing adoption of renewables, which are projected to account for nearly  40\% of global electricity output by 2027 \cite{oecdRenewables2022Analysis2022}.
Following the November 2006 incident -- in which a planned high-voltage line shutdown in Northern Germany caused overloading, splitting the European electricity grid into three zones and resulting in unforeseen regional power imbalances and blackouts affecting millions -- a gradual evolution towards regional system operators has also been recommended by the European Commission, emphasizing the need for more granular, decentralized energy forecasts
\cite{BlackoutNovember2006}.

The decentralization complicates  
the forecasting process, 
as numerous small-scale producers and consumers contribute to the overall energy landscape, posing new challenges that complicate accurate forecasting. 
These regional and decentralized stakeholders, 
ranging from small-scale local energy providers 
-- often managed as part of virtual power plants (VPPs) \cite{gaoReviewVirtualPower2024}-- 
to large building complexes
must accurately predict their energy production and consumption to participate effectively in energy markets,
avoid grid curtailment, and support overall grid stability.
Moreover, transmission system operators are increasingly  responsible
for managing peak capacity and providing balancing services, 
while distributed system operators must handle voltage support
\cite{ExecutiveSummaryUnlocking}.
In this complex environment, despite extensive research efforts
to forecast energy usage in electrical grids, these
operators continue to encounter significant outliers during unexpected scenarios
\cite{salehDataMiningBased2016}
as they are challenged by stochastic resource characteristics
\cite{gaoReviewVirtualPower2024}.

Load demand uncertainty is a common challenge for local energy providers, arising not only from seasonal fluctuations, but also from factors such as consumer behavior, current economic performance, production activities, and emergency or other events \cite{yuUncertaintiesVirtualPower2019}.
A significant   contributing  factor at a decentralized demand level is that 
smaller consumers often exhibit greater  variability and unpredictability in their energy consumption patterns due to diverse and individualized usage behaviors, thereby increasing load demand uncertainty
\cite{plaumAggregatedDemandsideEnergy2022, richardsonDomesticElectricityUse2010}.
This increased variability makes it more difficult to achieve high forecasting accuracy, 
as conventional forecasting models may struggle to capture the specific dynamics of small-scale energy interactions.
Traditional  forecasting approaches  typically emphasize regular fluctuations and established consumption trends, which may not adequately account for atypical events or unforeseen behavioral changes. As a result, these models struggle to accurately predict energy usage during unusual, non-recurring, or atypical  circumstances, leading to substantial forecasting errors.

To address the higher variability and unpredictability, researchers have enhanced load forecasting models by incorporating 
additional contextual information, with a strong focus on environmental forecasting data 
\cite{hongEnergyForecastingReview2020}.
Power grid operators integrating renewable energy sources utilize meteorological forecasts 
to improve load forecasting accuracy
\cite{markovicsComparisonMachineLearning2022}. 
For wind power forecasting, predictions on  wind speed and direction at different altitudes, 
temperature, atmospheric pressure, 
and humidity  are typically used as contextual information
\cite{lopezShorttermWindSpeed2022}. 
Photovoltaic power forecasting relies on forecasts of  
cloud coverage and imaging from cameras and satellites 
\cite{siPhotovoltaicPowerForecast2021, markovicsComparisonMachineLearning2022}.
Recent advances in numerical weather prediction 
have been described as a quiet revolution 
\cite{sweeneyFutureForecastingRenewable2020} 
and its adoption for renewable energy forecasting 
have significantly improved the accuracy of load forecasts. 
However,  the integration of additional  contextual information about the expected future 
-- such as anticipated user behavior and scheduled events -- 
in energy forecasting for industrial and residential consumers 
remains less advanced.
Although several industrial consumer case studies have incorporated operational timelines
-- such as factory operations 
\cite{bracaleShorttermIndustrialReactive2019},
production schedules 
\cite{zhuDayaheadIndustrialLoad2023}, 
and company holidays 
\cite{berkProbabilisticForecastingIndustrial2018} -- 
into load forecasting, 
research on the integration of extensive application-specific data remains limited. 
While individual studies have explored innovative approaches, 
such as one focusing on electrified cranes \cite{alasaliDayaheadIndustrialLoad2018},
a broader, systematic investigation of industrial energy forecasting remains limited.
Moreover, it has been recognized  that opportunities in the industrial sector
have not been comprehensively reviewed, as noted by Hong et al.  
\cite{hongEnergyForecastingReview2020}.

However, the digitalization of numerous industrial sectors and the widespread adoption of the Internet of Things (IoT) in both industrial and consumer applications have significantly increased the availability of information relevant to the forecasting task
\cite{tengRecentAdvancesIndustrial2021}.
This increase  in accessible  data  plays a crucial role in grid management 
\cite{jacobsonDistributedIntelligenceCritical2019}. Driven  by advancements in information and communication technology and 
improved  communication between consumers and utilities,
this data is now being  collected centrally
\cite{hossainApplicationBigData2019,
barja-martinezArtificialIntelligenceTechniques2021,
javaidEnergyEfficientIntegration2018}.

Within this data, certain  events, particularly those previously causing significant outliers may be anticipated  or planned, 
such as scheduled maintenance, major events like sports gatherings, 
or known shifts in energy consumption patterns
\cite{foseEmpoweringDistributionSystem2024}.
For example, large industrial electricity consumers, 
such as those in the railway sector
\cite{zhangJointOptimizationTrain2020, 
haehnFreightTrainScheduling2020,
heilRailwayCrewScheduling2020}, 
and the steel industry \cite{karelahtiLargeScaleProduction2011}, 
often meticulously plan their operations in advance.
We refer to such information 
as the "\textit{expected future}" \cite{abbottUnderstandingManagingUnknown2005}, which encompasses aggregated consumer data derived  from planning or other forecasting models. 
This  includes production plans, planned occupation, information on large gatherings,
vacation schedules, and timetables. 
This \textit{expected future} data is presumed  to represent recurring patterns  within the usual operating regime
-- describing regularly reoccurring scenarios and situations whose impacts  have been captured by past observations --
and are therefore highly likely to occur as planned or anticipated.

Despite its potential, the integration of \textit{expected future} information into forecasting models
remains an under-explored resource for improving the prediction 
of challenging load profiles
\cite{hongEnergyForecastingReview2020, zhaoReviewPredictionBuilding2012a}. 
Outliers in these load profiles are frequently attributed to random disturbances
\cite{foseEmpoweringDistributionSystem2024, mamlookFuzzyInferenceModel2009},
even though comprehensive data often exists that could help identify their true causes.
While exogenous variables are typically integrated into forecasting models, 
they are often incorporated in an autoregressive manner and 
cannot be integrated over the entire forecasting horizon.
As we will demonstrate, this approach leads to inferior performance compared to more comprehensive integration strategies
\cite{christenExogenousDataLoad2020}.
In this study, we propose a novel approach to energy forecasting that leverages the  complete \textit{expected future} information. 
Specifically, we conceptualize the forecasting task as a combined forecasting and regression problem, utilizing  
not only  historical data but also the complete \textit{expected future} information on exogenous inputs.
We propose a  transformer-model-based methodology  designed to effectively incorporate dynamic and planning-related  \textit{expected future} information, 
such as user behavior schedules and scheduled events. 
This integrated approach  improves forecast accuracy and reduces the occurrence  of large outlier predictions.
\cite{wangTimeXerEmpoweringTransformers2024}.

The proposed load forecasting framework shows its efficacy in predicting nationwide railway energy consumption integrating  
timetable data, with an auxiliary case study on building energy usage forecasting based on planned office occupancy. 
Our results demonstrate that integrating   \textit{expected future} contextual information 
into transformer models  significantly improves forecasting accuracy. This integration achieves an average reduction of \textbf{26.6\%} 
in mean absolute error for the best-performing  railway energy consumption forecasting model
and a substantial \textbf{56.3\%} reduction for the top  building energy forecasting model. 
Furthermore, our framework  reduces the number of outlier predictions in both case studies, addressing a critical challenge  in energy forecasting.

We demonstrate, that the integration of the \textit{expected future} information reduced the amount of significant outliers for the best performing models
in the \textit{Railway} forecasting scenario by \textbf{87.8\%} and for the \textit{Building Energy} case by \textbf{93.0\%}.
In contrast, other state-of-the-art forecasting methods failed to show significant improvements, 
highlighting the superior performance of our approach in leveraging "expected future" contextual data for enhanced energy forecasting.
By effectively integrating \textit{expected future } contextual information, 
our method not only enhances forecasting accuracy and reduces the numbers of outliers but also contributes to more resilient and cost-effective energy systems.

\section{Integration of the Expected Future across  Electrical Energy Domains}

Our research demonstrates that contextual expected future information is essential for accounting for variability in load predictions, particularly in capturing outlier events. While previous studies have integrated weather forecasts -- a type of expected future information -- into load predictions, they have often neglected other readily available data, such as planning information and anticipated rare events.   
To evaluate the versatility and effectiveness of the contextually enhanced transformer models, as well as the benefits of integrating diverse types of future contextual information, we present two different case studies.
Our primary case study focuses on day-ahead grid load forecasting for the Swiss RTN integrating historical data with expected future contextual information, such as passenger and freight timetables,  and operational plans. 
Given that the Swiss RTN is part of the European RTN  -- the largest interconnected 15 kV 16.7 Hz system spanning Germany, Austria and Switzerland -- it serves as an exemplary showcase for regional production and demand balancing.
This application demonstrates the importance of integrating and aggregating planning information from diverse sources to improve prediction accuracy. Additionally, we evaluate the impact of different levels of geographical information aggregation, demonstrating how spatial granularity can influence forecasting performance.

In the auxiliary case study, we apply our framework to forecast energy consumption in building systems by utilizing contextual variables such as dynamic occupancy profiles. while buildings are significant  energy consumers with  potential for energy efficiency optimization based on accurate load forecasts, this case primarily serves to demonstrate  the broader applicability of our approach rather than introducing a significant novelty, as similar evaluations incorporating weather and occupancy information have been conducted in the past
\cite{kusterElectricalLoadForecasting2017}. 
Nonetheless, the results underscore  the value  of integrating future contextual information about building indoor conditions for load forecasting in the case of a medium-sized office building.

The framework's ability to generalize across these two distinct domains domains highlights  its broad applicability and potential to  improve load forecasting accuracy, ultimately  supporting more robust and data-driven energy management strategies.

\subsection{Primary Case Study: Forecasting Dynamics in Railway Traction Networks}

Railway networks and their dedicated power grids for electrified railways are critical components of national and regional transportation infrastructure, enabling the efficient and reliable movement of passengers and freight while supporting the shift toward sustainable energy systems. 
Railway networks are uniquely suited to leverage expected future information, as passenger timetables are typically well-planned, long-term, and recurring, while the dynamic scheduling of freight trains demands adaptable and responsive forecasting.
Accurate load forecasting for these networks is essential to optimize energy consumption, enhance operational efficiency, and ensure the stability and sustainability of vital transportation services.

The Swiss Federal Railways (SBB) operates  a 
dedicated RTN
at 132kV~/~15kV, single phase,
specifically designed to power the rolling stock 
of the national railway network, 
which facilitates over 8000  train journeys daily. 
The railway traction grid is a substantial entity, encompassing over 1,800 kilometers of transmission lines, 70 substations, and a robust infrastructure of 13 power plants and converter stations, making use of pumped-storage hydroelectrical power plants to address the challenge of balancing supply and demand
\cite{BahnUndHaushaltsstrom}.
The RTN operates at a unique 16.7 Hz frequency, distinct from the standard 50 Hz used in the conventional consumer grid.
Its power supply is well supported by dedicated power plants and integrated with the main grid through  11 frequency converters, ensuring  a stable and well-regulated energy environment. These controlled and well-defined operational conditions  make the RTN an ideal case study for investigating the impact of contextual information on forecasting accuracy and performance.

The primary objective  is to accurately predict  the next day’s grid load on an hourly basis
to support day-ahead planning for energy production and trading in energy markets.
The RTN is a complex and expansive infrastructure
that operates according to well-defined, recurring operational patterns
meticulously scheduled  through precise timetables. 
It manages both passenger transportation  and detailed freight train schedules, both of which are essential  for forecasting  the subsequent  day's energy consumption.
However, forecasting in this complex system is challenging, as the RTN not only powers electric trains but also supports a variety of signaling systems, station infrastructure, and auxiliary equipment.
Therefore, grid load forecasting for the RTN is influenced 
not only by historical trends and periodicity
but also by \textit{future contextual information}.
Key variables derived from centralized planning include 
the allocation of railway rolling stock  (encompassing  both long-term strategic planning and short-term scheduling of specific rail vehicles for particular journeys or services), and 
planned long-term construction projects
\cite{boschPrognosenLeistungsbedarfsVolatiler2017}. 
 Additionally, we incorporate  weather forecasts as additional expected future variables, recognizing their significant impact on energy usage within the RTN.

The case study for this research consists of  two multi-year datasets,
\textit{Railway} (2018-2023) and \textit{Railway-agg} (2020-2023), recorded at different aggregation levels with different number of contextual information, both
captured at hourly resolution.
These datasets contain grid load along with 
relevant covariates aimed at  enhancing forecasting accuracy 
\cite{boschPrognosenLeistungsbedarfsVolatiler2017}.
The recorded covariates include temperature recordings,  
gross tonne-kilometers moved (which combines tonnage and mileage), 
and train counts, all derived from the timetable
of the Swiss national railway traction network. These are used  
to forecast the next day's nationwide railway energy consumption at hourly basis. Additional details can be found in Section \ref{sec:datasets}.
The data will be made publicly available following  the acceptance of this paper.

\begin{figure}[h]
\centering

\makebox[\textwidth][c]{%
    \includegraphics[width=1.3\textwidth]{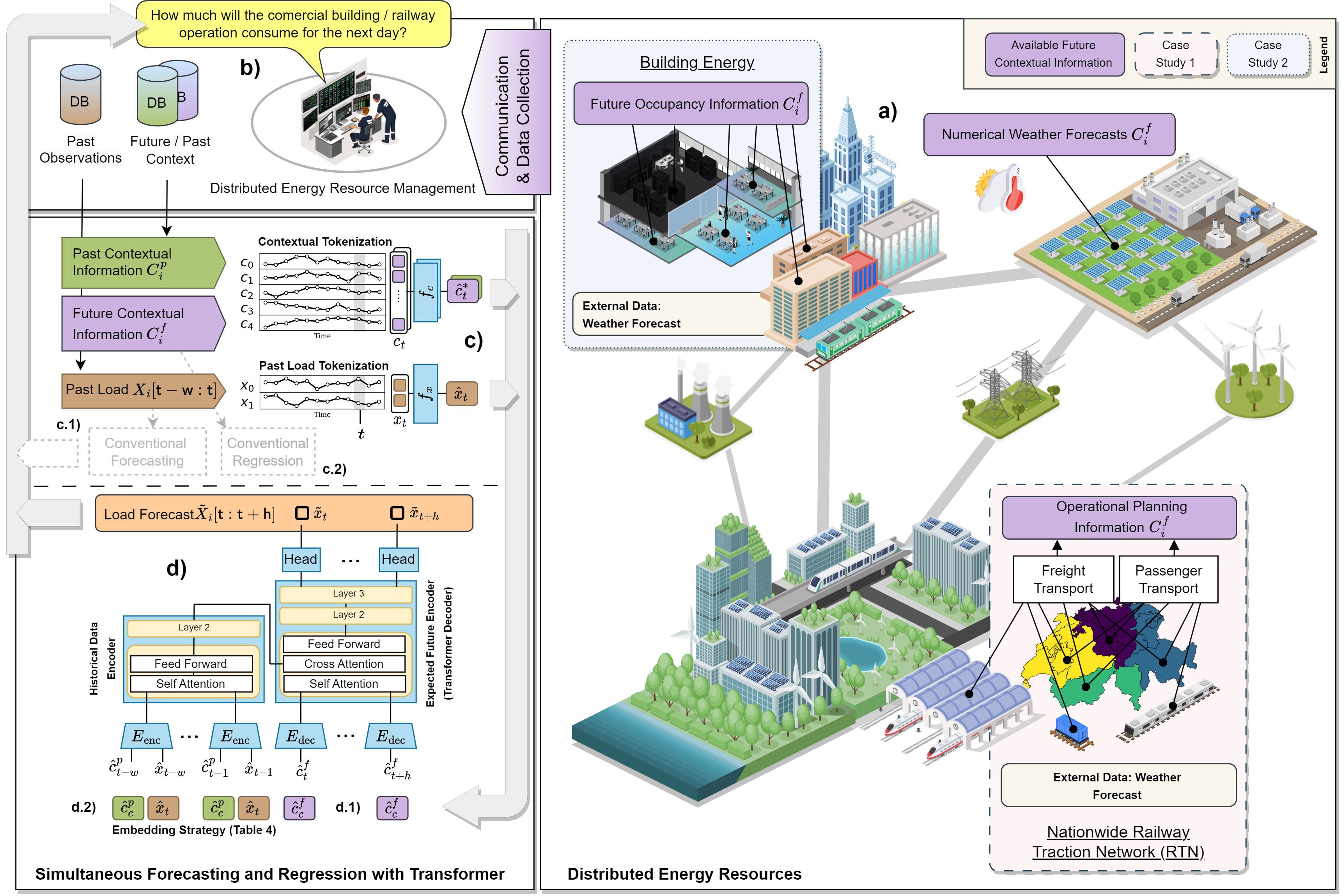}
}
\caption{
Illustration of the proposed load forecasting framework with contextually enhanced transformer models, highlighting the case studies focused on the Swiss national railway traction network (\textit{Railway} and \textit{Railway-Agg} dataset) and load forecasting for buildings (\textit{Building Energy} dataset) in Panel \textbf{a}.
Panel \textbf{b} displays the collection of "\textit{expected future}" data, including future occupancy information from building management,
numeric weather forecasts,
timetables, 
schedules and gross ton-kilometers (GTKM) estimates derived 
from the operational planning of the railway operator.
Traditionally, methods such as  pure timeseries forecasting \textbf{c.1} and 
regression models \textbf{c.2} are employed for load forecasting.
Our proposed approach introduces the use of transformer architectures to learn a unified representation
of the \textit{time series regression task} (\textbf{d}).
To efficiently integrate both past and future information for this task, 
we propose dividing  the input data at the current time point $t$ (the present) and to tokenize the segments individually (\textbf{c}). We then apply distinct  embedding strategies for past data (\textbf{d.2}) and 
\textit{future contextual information} (\textbf{d.1}) in our contextually enhanced transformers. 
}
\label{fig:visual_abstract}
\end{figure}

\subsection{Auxiliary Case Study: Forecasting Dynamics in Building Energy Systems}

Load forecasting for buildings is critical, as they account for approximately 40\% of total energy consumption globally  \cite{hongEnergyForecastingReview2020, zhaoReviewPredictionBuilding2012a}, 
with about one-third of global energy consumption attributed to building operation \cite{iea_buildings}. The increasing electrification of heating and mobility services further elevates  the importance of building power demand for their smooth integration into local distribution networks. 
Accurate load forecasts support energy cost management, improve operational efficiency by optimizing Heating, Ventilation, and Air Conditioning  (HVAC) systems, 
and enable better integration with local renewable energy sources, particularly for larger buildings such as offices. 
Buildings are also well-suited for incorporating future contextual information, 
as occupant behavior and usage patterns -- key drivers of energy demand -- 
can provide valuable insights to enhance forecasting accuracy and align energy consumption with expected usage.

To demonstrate  the generalizability of our forecasting framework, the auxiliary case study on \textit{Building Energy} forecasting  focuses on an individual medium-sized office building in the United States, a subset of the dataset \textit{AlphaBuilding}, registered with the U.S. Department of Energy’s Open Energy Data Initiative (OEDI) \cite{li_synthetic_2021}. The dataset represents 70\% of U.S. commercial buildings and is widely used within the building energy community. The dataset is generated using the building simulation software EnergyPlus \cite{energyplus}, which models the office building by incorporating building characteristics, HVAC system specifications, and weather data. \\
The target building consists of three floors with a total floor area of 4'890 square meters, constructed in a vintage style according to ASHRAE 90.1-2013 standards \cite{ashrae90.1-2013}.
Furthermore, the building encompasses a HVAC system consisting of air handling units and its load profiles are simulated with the help of weather data, generated occupancy profiles as well as lighting and miscellaneous electric load schedules.
The dataset includes weather data, occupancy aggregated per floor, and total building energy consumption. The total building energy serves as the target variable, while weather data and occupancy represent the context variables. Additional details can be found in Section \ref{sec:datasets}.

This building energy case study is considered auxiliary, as similar studies leveraging weather and occupancy information for energy forecasting have been conducted in the past. While it serves as a validation of our framework’s ability to integrate future contextual information, the primary contribution of our work lies in enhancing load forecasting for large-scale, dynamic energy systems such as railway networks, where contextual information has not been effectively utilized in prior approaches. As such, the railway energy forecasting case study remains the primary focus, while the building energy study illustrates the broader applicability of our method across different domains.

\subsection{Comparative Evaluation of Contextually Enhanced Transformers Across Architectures and Models}
We evaluate  our contextually enhanced transformers
across  three forecasting tasks from two case studies utilizing multivariate time-series datasets enriched with future contextual information. 
Specifically, the  \textit{Railway} and \textit{Railway-agg} datasets
include  extensive \textit{Future Contextual Information} (FCI)
 from scheduling and operational planning, while the \textit{Building Energy} dataset includes  occupancy data sourced from planning or IOT sensor forecasts.
To demonstrate  the flexibility and architecture-independence  of our  contextually enhanced transformer
approach, detailed  in Section \ref{sec:efficient_integration_of_expected_future}, we apply  it to   three distinct 
encoder-decoder transformer
architectures:
the Contextually Enhanced Crossformer (CE-CF), 
the Contextually Enhanced  Time-Series Transformer with Informer embedding (CE-TST);
and the Contextually Enhanced Spacetimeformer (CE-STF).
For each transformer architecture, we compare our contextually enhanced variants against their standard,
unaltered variants  to establish baseline performance. 
Additionally, 
we evaluate  the efficacy of recent state-of-the-art (SOTA) multi-step linear models, DLinear,  
and its extension TiDE.
We also include traditional machine learning models as additional baselines to enhance our evaluation. These models include BiLSTM, Multilayer Perceptron (MLP)
a conventional linear regression method (EUB), K-Nearest Neighbor (KNN) Regression
and the gradient-boosting framework CatBoost. Further, we evaluate traditional time series models, namely the weekly seasonal naive method and AutoSARIMAX. Incorporating these models offers  a wide  range  of comparative insights, enabling a more thorough  evaluation of our methodology.
Additionally, we explore  the inverted Transformer (iTransformer) strategy on the STF, 
as proposed in \cite{liuITransformerInvertedTransformers2023} and PatchTST.
The models are referenced in Section \ref{sec:baseline_methods}.

\begin{figure}[h]
    \centering

\begin{subfigure}[t]{0.48\textwidth}
\centering
\includegraphics[width=\textwidth]{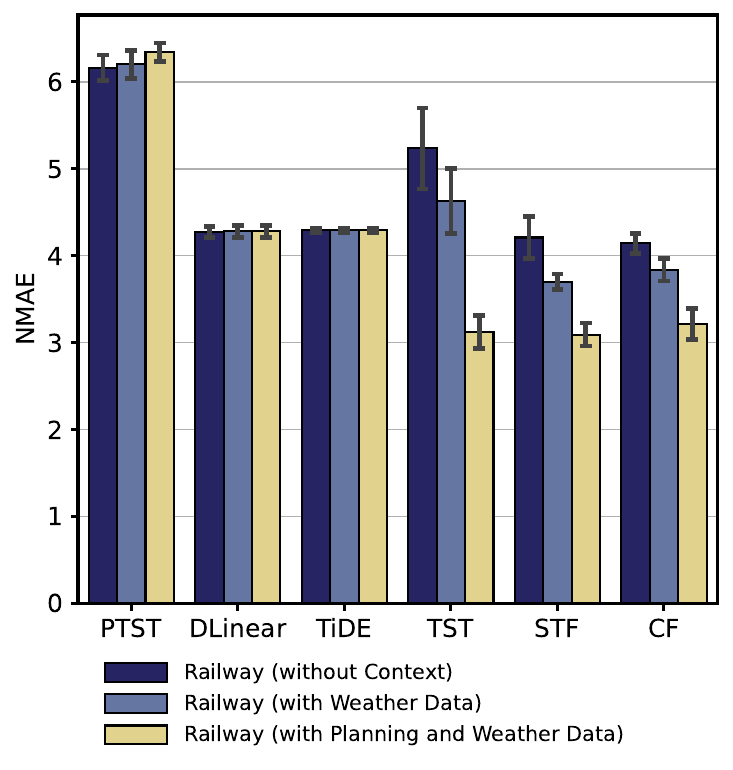}
\caption{
NMAE performance for the \textit{Railway} test set.
}
\end{subfigure}
\hfill
\begin{subfigure}[t]{0.48\textwidth}
\centering
\includegraphics[width=\textwidth]{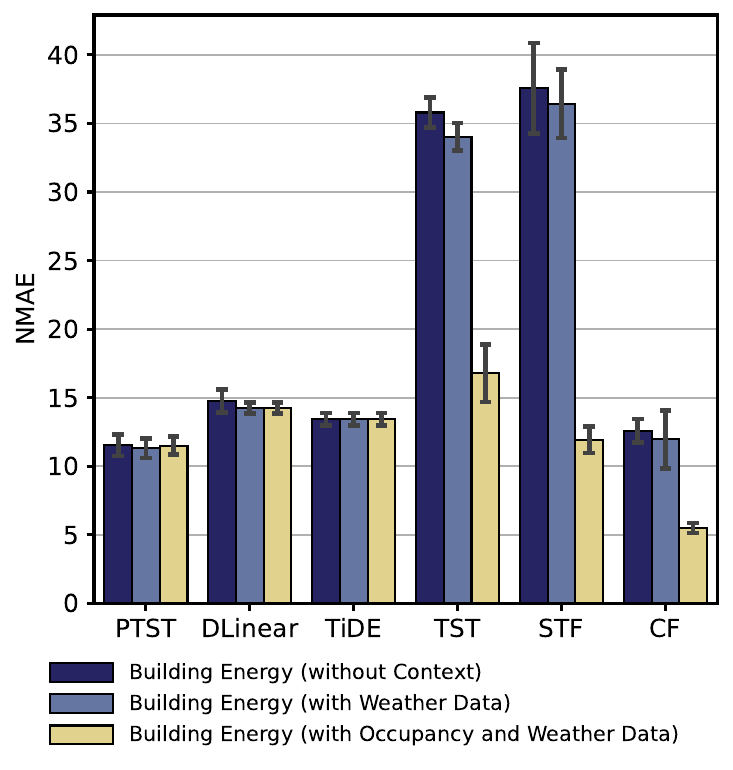}
\caption{  
NMAE performance for the \textit{Building Energy} test set.
}
\label{fig:overall_model_performance_alphabuilding}
\end{subfigure}

\caption{Normalized Mean Absolute Error (NMAE) in normalized megawatts with and without the addition of FCI 
on the \textit{Railway} and \textit{Building Energy} dataset.
We list all contextually enhanced transformer models:
enhanced Crossformer (CF),
enhanced Spacetimeformer (STF) and 
enhanced Timeseries Transformer (TST),
PatchTST (PTST),
and multi-step linear models (DLinear and TiDE) included in our evaluations.
}
\label{fig:average_overall_improvements_fci}
\end{figure}

\textbf{The effectiveness of Future Contextual Information (FCI) for load forecasting:}
Incorporating FCI into various transformer architectures significantly enhances their predictive accuracy, resulting in substantial improvements in forecasts for both case studies. (Table \ref{tab:main_text_baseline_methods}).
Specifically, for the \textit{Railway} dataset, 
integrating future planning data reduces the  NMAE of the best-performing model,  Contextually Enhanced Spacetimeformer,  by \textbf{16.5\%}. 
In building energy forecasting, 
the inclusion of future occupancy information  decreases the NMAE of the best-performing model, Contextually Enhanced Crossformer,  by  \textbf{54.0\%}, 
resulting in a total reduction of the NMAE \textbf{53.8\%} 
across both case studies and 
a reduction of the standard deviation across all trainings 
by \textbf{26.1\%}.  
These results demonstrate  that while contextually enhanced transformers effectively leverage contextual information to improve prediction accuracy, whereas 
the addition of FCI does not yield similar benefits for TiDE and DLinear and PatchTST models (see Figure \ref{fig:average_overall_improvements_fci} ).
Notably, despite  TiDE's 
feature encoder being specifically designed  for dynamic covariates from both past and future contexts,  no improvement 
in forecasting accuracy was observed.

The combined impact of weather information and additional FCI is further evidenced in specific models for the \textit{Railway} and (\textit{Building Energy}) dataset:

\begin{itemize}
    \setlength{\itemsep}{0pt} 
    \setlength{\parskip}{0pt} 
    \setlength{\topsep}{0pt} 
    \setlength{\partopsep}{0pt} 
\item The Crossformer model sees a \textbf{49.9\%} (\textbf{56.3\%}) reduction in NMAE.  
\item The Spacetimeformer achieves a   \textbf{26.6\%} (\textbf{68.2\%})  decrease.  
\item The Time-series Transformer experiences a  \textbf{40.3\%} (\textbf{53.1\%})  reduction. 
\end{itemize}

These improvements highlight  the substantial  advantage  of integrating  \textit{expected future} contexts into the forecasting process, as further detailed in the ablation experiments in Table \ref{tab:ablations_railway} and Table \ref{tab:ablations_building} of the supplementary material.

\textbf{Day-ahead load forecasting in railway traction networks:}
For railway grid forecasting, we categorize our analysis into two  setups  based on data  from two distinct operational planning models. 
The \textit{Railway-agg} dataset aggregates national-level data,
while the \textit{Railway} dataset offers a more granular perspective with detailed data from four separate geographic regions.
Both datasets are utilized to forecast the total grid load. 
Our analyses across   all evaluated datasets clearly demonstrate that our contextually enhanced transformers consistently outperform 
all baseline methods, including all versions of the transformers without the proposed extension.
Detailed performance metrics of the forecasting models on the 
\textit{Railway} and \textit{Railway-agg} datasets are presented in 
Table \ref{tab:main_text_baseline_methods}.
The contextually enhanced CE-STF stands out as the top-performing model across  both datasets, notably achieving the lowest NMAE of \textbf{3.09}  (megawatts normalized by the mean of the data set) and NRMSE of \textbf{3.95}.
In contrast, the performance of the state of the art multi-step linear models such as DLinear and  TiDE lags behind, with NMAEs of \textbf{4.28}  and \textbf{4.29} , respectively. We identified conventional linear regression (\textbf{3.85} NMAE) and CatBoost (\textbf{3.67} NMAE) as the best performing baseline models. However, CE-STF surpasses CatBoost, achieving a \textbf{15.8\%} improvement in MAE.
CE-STF also outperforms the existing conventional linear regression model (EUB) that is currently in production.

\textbf{Day-ahead load forecasting in buildings:} For building energy forecasting, the contextually enhanced CE-CF outperforms all other benchmark models, achieving a NMAE of \textbf{5.50}  (see Table \ref{tab:main_text_baseline_methods}). Additionally, CE-CF shows the lowest variability across  different training seeds, indicating robust and consistent performance. Notably, the weekly seasonal naive method is the second best performing model, benefiting from  the regular patterns inherent in the simulated building energy data set. However, the  normalized error is  twice that of CE-CF. Following closely are CE-STF and Catboost, which demonstrate comparable performance to the weekly naive method. In contrast, other benchmark models demonstrate substantially worse error metrics. Linear Regression, KNN Regression, MLP and BiLSTM display a similar performance around a NMAE of \textbf{21} . Even though the model architectures are considerably different, all models struggle to accurately predict the load for low-load days. Further, AutoSARIMAX exhibits significantly higher error metrics with a NMAE of \textbf{56.43} . This under-performance may result from the suboptimal selection of SARIMAX orders,  leading to underestimation of  high-load times during the day.

\textbf{Is load forecasting a regression or forecasting problem?}
Our hypothesis suggests  that forecasting in scenarios where rich future contextual information is available
can be effectively approached  through two primary methods:  
historical data analysis and regression techniques that integrate anticipated future contexts
\cite{nabaviDeepLearningEnergy2021}.
To further examine the impact of contextual information, we have conducted  
several ablation studies on the railway grid forecasting case study on the STF, the top-performing contextually enhanced transformer model in our previous experiments (supplementary Table \ref{tab:ablations_railway}).
In our ablation study on the \textit{Railway} dataset, we examine the impact of incorporating future contextual information, distinct from merely integrating weather forecasts, on forecasting accuracy. The results reveal  that significant improvements in  forecasting accuracy can be achieved by incorporating both: past time series data (w/o Enc Load) and contextual information.
Specifically, adding Future Contextual Information (FCI) to the decoder results in  a \textbf{26.6 \%} 
reduction in mean NMAE, while  integrating past time series data into the encoder leads to  a \textbf{29.1 \%} reduction.  
Interestingly, the addition of past contextual information alone yields only a minor
improvement of \textbf{3.1 \%} 
in mean NMAE. This suggests that much of  of the value of contextual information may already be captured within the past time series itself.
These findings highlight  the  dual nature of the load forecasting problem, presenting  challenges typical of both regression and forecasting tasks.
In an ablation study  assessing the duration of past contextual information on the railway dataset, we observe  that
transformer models do not experience performance improvements when provided with longer contextual inputs
on the \textit{Railway} dataset.
Increasing the context length from one day to eight days (\(w=192\)), thereby 
providing the relationships from the previous day and the target load profile for the previous week as input to the model, surprisingly  
degrades forecasting performance by \textbf{24.8\%}. 
In contrast, multi-step linear models
TiDE and DLinear require a considerably longer context sequence of one month (\(w=672\)) for effective forecasting, as shown 
in the supplementary Table \ref{tab:ablations_railway}. This
underscores the dependency of linear models 
on past trends and periodicity. 
However, the performance of these models still lags behind that 
 of contextually enhanced transformers, highlighting  the critical need for efficient  integration of FCI.
 
In case of the \textit{Building Energy} case study, our analysis reveals that incorporating future context information leads to substantial improvements in forecasting accuracy for several models (see Table \ref{tab:ablations_building}). Notably, Crossformer and STF achieve an NMAE reduction of \textbf{56.30 \%} resp. \textbf{68.24 \%}, demonstrating the significant benefits of leveraging the contextual information. Additionally, Linear Regression, AutoSARIMAX and Catboost demonstrate substantial reductions in forecasting errors when future context is integrated, highlighting their ability to effectively utilize supplementary information. In contrast, models such as DLinear or TiDE show minimal to no improvement, indicating challenges integrating and benefiting from the future contextual data. The table also shows the isolated effect of including future occupancy information for Transformer models and DLinear. We see that most of the performance stems from the past time series (w/o FCI), and only adding weather improves the error metrics marginally (w/o OCC). However, additionally considering future occupancy information leads to the substantial improvement in NMAE of \textbf{54.00 \%} in CE-CF and \textbf{67.25 \%} for STF. Conversely, DLinear does not improve performance in this case. These insights again underscore  the  dual nature of the load forecasting problem. 

In both case studies (supplementary Table \ref{tab:ablations_railway} and \ref{tab:ablations_building}), we find that prior efforts to integrate exogenous variables into autoregressive models, such as SARIMAX (Seasonal AutoRegressive Integrated Moving Average with eXogenous variables), were outperformed by contextually enhanced transformer models.

\begin{figure}[h]
\centering
\begin{subfigure}[t]{0.48\textwidth}
\centering
\includegraphics[height=\textwidth]{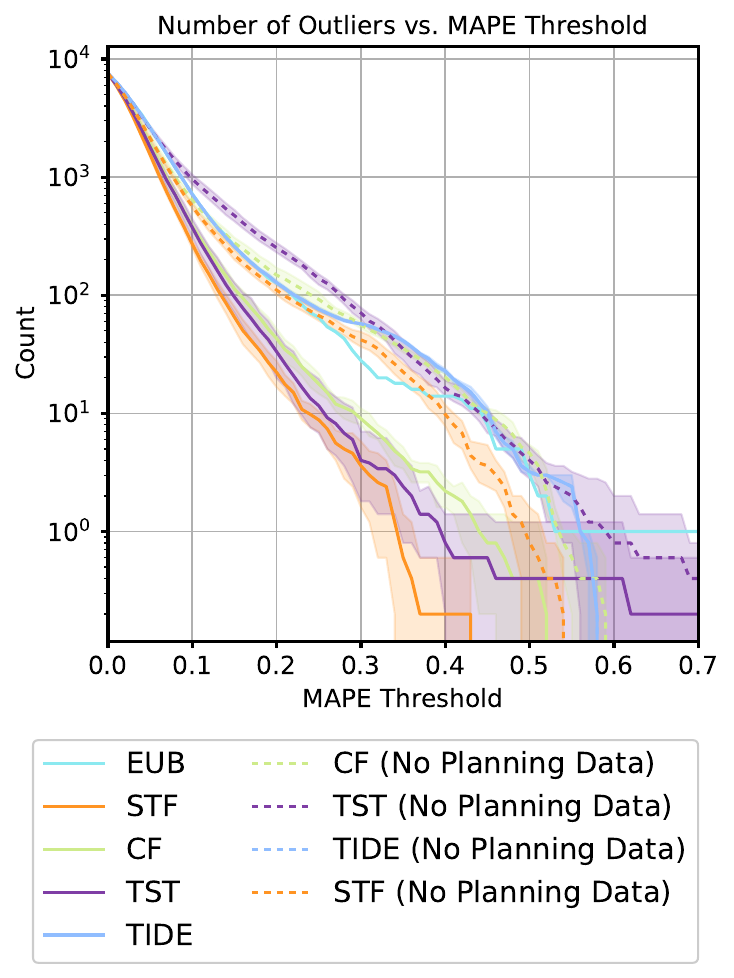}
\caption{
Outlier counts by forecasting model plotted against the MAPE threshold for the \textit{Railway} test set.
}
\label{fig:mape_outliers_railway}
\end{subfigure}
\hfill
\begin{subfigure}[t]{0.48\textwidth}
\centering
\includegraphics[height=\textwidth]{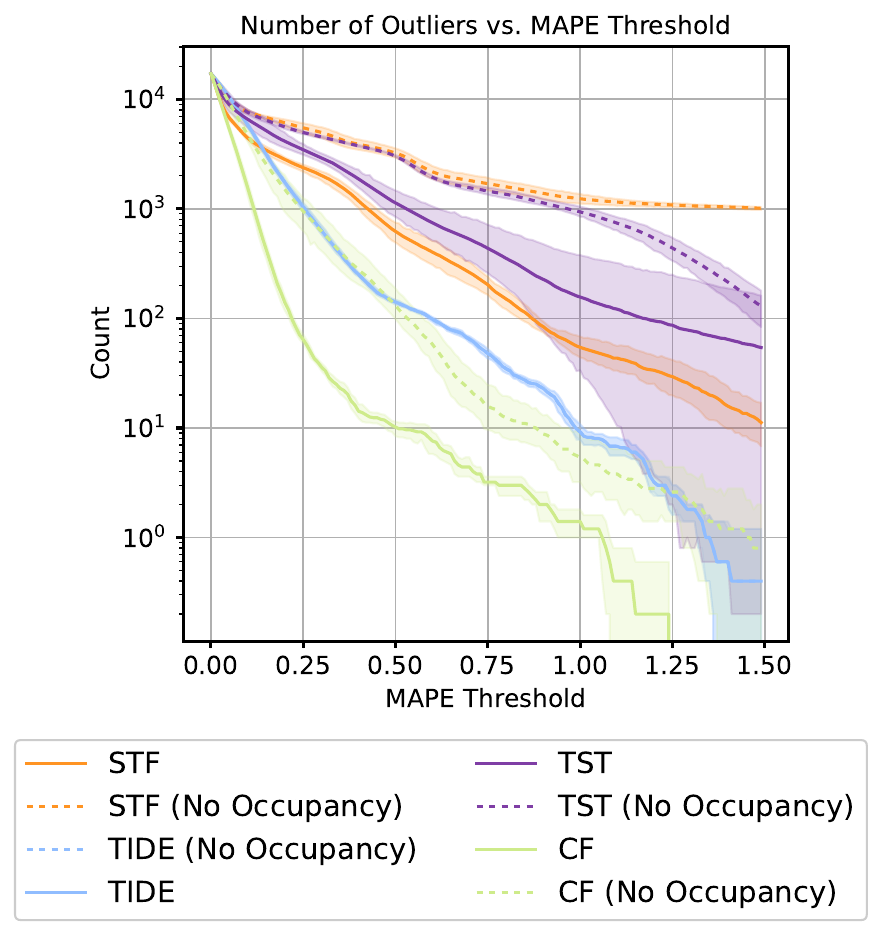}
\caption{  
Outlier counts by forecasting model plotted against the MAPE threshold for the \textit{Building Energy} test set.
}
\label{fig:mape_outliers_building}
\end{subfigure}
\caption{
Comparison of the robustness
of contextually enhanced transformer models: 
Crossformer (CF),
Spacetimeformer (STF) and 
Timeseries Transformer (TST)
trained and evaluated on the \textit{Railway} dataset in a) and on the \textit{Building Energy} dataset in b).
The linear regression model (EUB), 
currently the best performing model in production at the data supplier, is also included for comparison in a).
Error bands illustrate  the variation across different training initializations.
}
\label{fig:robustness_comparison}
\end{figure}

\begin{figure}[h]
\centering
\includegraphics[width=\linewidth]{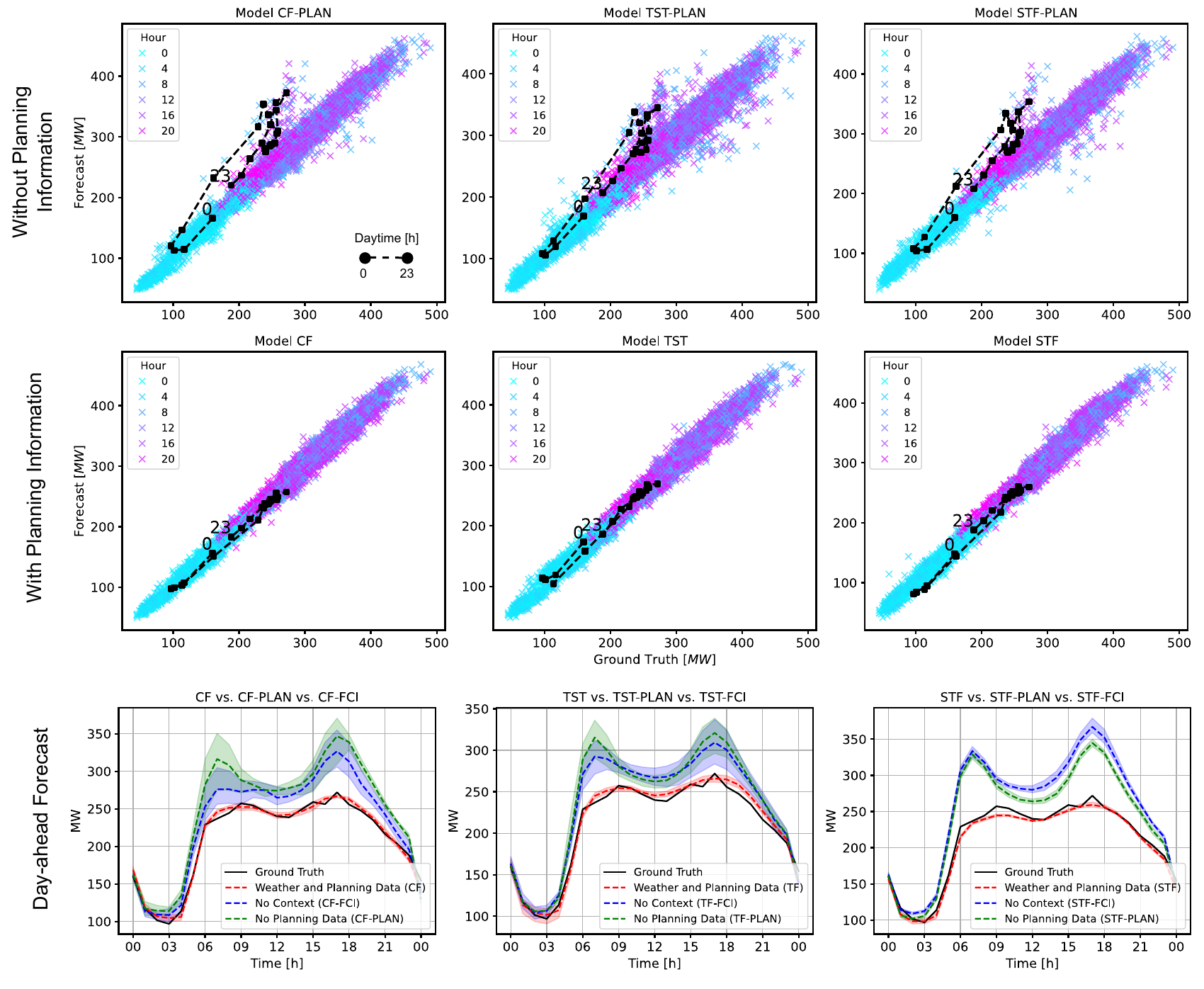}

\caption{
\textbf{Model Performance Case Study: Swiss National Holiday (August 1, 2023)}
This detailed study focuses on the Swiss National Holiday event 
in the \textit{Railway} dataset.
For the individual contextually enhanced transformer Crossformer (CF),
Spacetimeformer (STF) and Timeseries Transformer (TST) 
we show scatter plots relating forecasted values to ground truth for the entire test set. In the first row we show the models without
planning information (-PLAN), in the second row, we include planning information.
We overlay the 24 time steps of August 1 in black.
Below, we present the predicted load curves for forecasts made with and without future contextual information. To highlight the impact of different data sources, we separately examine planning data and weather data in the forecast plot, illustrating the substantial benefits of integrating planning data alongside weather data.
Error bands are included to represent the variability across multiple training runs.
}
\label{fig:case_study_august_1}
\end{figure}

\textbf{Analyzing forecasting outliers:}
In addition to improved average accuracy, our proposed approach significantly reduces both the frequency and magnitude of large outliers by leveraging  contextual information as displayed in Figure \ref{fig:robustness_comparison}. 
These difficult-to-compensate outliers are typically a major concern for power grid operators  
as they can pose a risk to grid stability or cause substantial financial losses due to the necessity of last-minute emergency purchases in the intra-day trading market. \\
In the \textit{Railway} case study, our analysis reveals that while  transformer models generally perform acceptably on average; however, they tend to produce  a higher number of outliers without the integration of FCI.
Furthermore, we analyze the Swiss National Holiday (August 1, 2023) from the Railway dataset -- a major outlier -- in Figure \ref{fig:case_study_august_1}, comparing forecasting performances of transformer models with and without planning and weather data. On that date, we showcase the impact of integrating contextual features to mitigate the outlier.
Overall, transformer models without FCI, exhibit   an average of \textbf{0.60\%} significant outliers exceeding a 30\% MAPE for the transformer models, which is reduced to \textbf{0.073\%} outliers when FCI is included.
Upon integrating  FCI,
all contextually enriched transformer models significantly outperform the linear regression model EUB (\textbf{0.37 \%}) in managing outliers.
This  analysis also highlights  the relative robustness of the different contextually enhanced transformers, 
as illustrated in Figure \ref{fig:robustness_comparison}. Although  CE-STF, CE-TST and CF demonstrate similar average performances in terms of NMAE, NRMSE, MAPE, and coefficient of determination (supplementary Figure \ref{fig:averaged_performance_comparison}), CE-STF enhanced with FCI exhibits
the lowest count of outliers and the smallest  maximum outlier magnitude, 
establishing it as the most robust model against outliers. \\
In the \textit{Building Energy} case study, we observe a comparable performance improvement among the contextually enhanced transformer models in terms of reducing outlier counts through the integration of future contextual information. CE-CF with contextual information consistently records  the lowest outlier counts across all MAPE threshold values. 
Particularly, CE-CF displays an average of \textbf{2.97 \%} significant
outliers exceeding a 30\% MAPE which is lowered to \textbf{0.2 \%} outliers when FCI is included. CF also reaches a zero count of outliers already at a MAPE threshold of \textbf{1.25}, whereas the exclusion of FCI raises the threshold to \textbf{1.92} MAPE. This pattern reinforces the advantage of incorporating future context in reducing forecasting outliers and establishes contextually enhanced Crossformer as the most robust model against outliers.
Moreover, the inclusion of future context universally helps to reduce the outlier counts across models.

\begin{figure}[h]
\centering
\includegraphics[width=\linewidth]{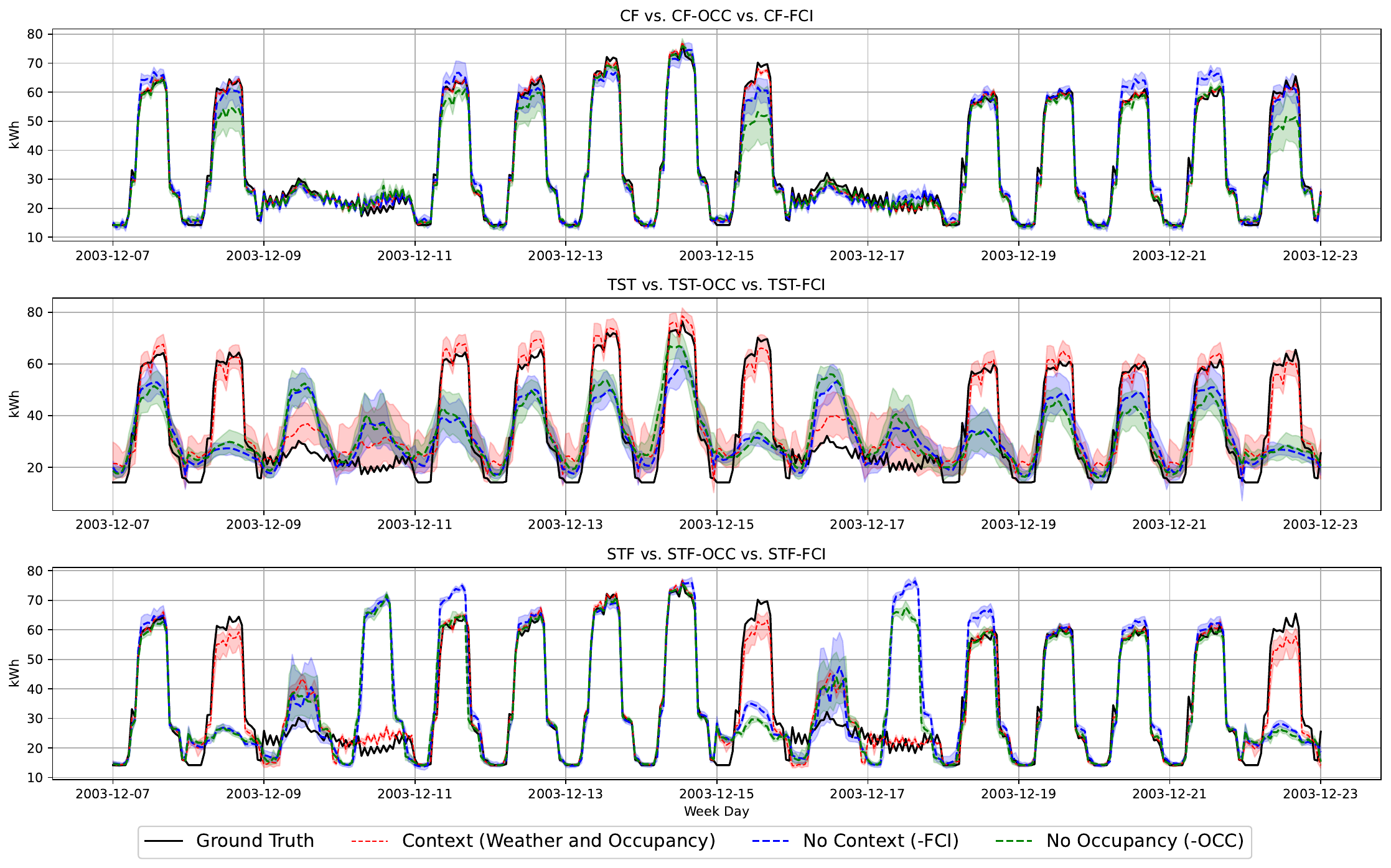}

\caption{
Model Performance case study for the \textit{Building Energy} dataset. We overlay 
the building energy profile with the day-ahead forecasts (48 time steps) of contextually enhanced transformer models (Crossformer (CF),
Spacetimeformer (STF) and Timeseries Transformer (TST)).
We plot forecasts with and without \textit{future contextual information}(-FCI).
To highlight the impact of different future context sources, we separately display the impact of removing occupancy data (-OCC) in the forecast plot.
Error bands show variation across training runs.
}
\label{fig:typical_load_profile_building_energy}
\end{figure}

\textbf{Discussion of Model-specific Effects:}
Figure \ref{fig:typical_load_profile_building_energy} compares day-ahead forecasts of transformer models with and without future contextual information. 
The \textit{Building Energy} dataset is generated using yearly occupancy profiles that change annually, an unrealistic decision made by the original authors of the simulated dataset. 
Consequently, because the test set covers exactly one year, the pattern shifts present in the test set are not encountered during training.
Figure \ref{fig:typical_load_profile_building_energy} reveals that individual days with shifted patterns lead to large variations in forecasting error, explaining the performance differences observed in Figure \ref{fig:overall_model_performance_alphabuilding}.
The excellent performance of CE-CF in this scenario highlights the model's reduced dependency on numerical time encodings.
On the contrary, supplementary Figure \ref{fig:performance_weekdays_alphabuilding} demonstrates that CE-STF performs well on days with training set overlap,
exhibiting comparable performance to CE-CF,
but its separate time encoding layers make the model more sensitive to out-of-time distribution shifts.
Furthermore, we found it intriguing that the low impact of future contextual information (FCI) in TiDE can be attributed to LayerNormalization. When LayerNormalization is disabled in the context encoder, the model becomes sensitive to FCI but fails to converge during training.

\begin{table}[h]
\centering
\caption{
Normalized Mean Absolute Error (NMAE) and Normalized Root Mean Squared Error (NRMSE) test set performance  for the \textit{Railway}, \textit{Railway-Agg} and \textit{Building Energy} data set.
All datasets share the load as the target variable for forecasting. 
However, they differ in the type and level of detail provided for co-variates and length
(Table \ref{tab:data_splits} lists the data splits). 
The \textit{Railway-agg} dataset offers a smaller set of aggregated co-variates, 
while the larger \textit{Railway-agg} dataset provides spatially localized co-variates specific to different regions. The \textit{Building Energy} dataset contains aggregated occupancy co-variates for the office building.
}
\label{tab:main_text_baseline_methods}
\begin{tabular}{lllllll}
\toprule

Data Set
& \multicolumn{2}{l}{Railway-Agg}
& \multicolumn{2}{l}{Railway}
& \multicolumn{2}{l}{Building Energy} \\
\cmidrule(l){1-1}  \cmidrule(l){2-3} \cmidrule(l){4-5} \cmidrule(l){6-7}

\# Future Covariates 
& \multicolumn{2}{l}{\textit{16}}
& \multicolumn{2}{l}{\textit{56}}
& \multicolumn{2}{l}{\textit{14}} \\
    
Trend Decomp.
& \multicolumn{2}{l}{\textit{yes}}
& \multicolumn{2}{l}{\textit{no}}
& \multicolumn{2}{l}{\textit{no}} \\
\cmidrule(l){1-1} \cmidrule(l){2-3} \cmidrule(l){4-5} \cmidrule(l){6-7}
Model / Metric & NMAE & NRMSE  & NMAE & NRMSE  & NMAE & NRMSE  \\
\cmidrule(l){1-1} \cmidrule(l){2-3} \cmidrule(l){4-5} \cmidrule(l){6-7}
Weekly Naive & \text{\small 10.57} \text{\tiny ±0.00} & \text{\small 15.08} \text{\tiny ±0.00} & \text{\small 6.67} \text{\tiny ±0.00} & \text{\small 9.49} \text{\tiny ±0.00} & \text{\small 10.76} \text{\tiny ±0.00} & \text{\small 21.79} \text{\tiny ±0.00} \\
KNN Regression& \text{\small 4.39} \text{\tiny ±0.00} & \text{\small 5.91} \text{\tiny ±0.00} & \text{\small 5.05} \text{\tiny ±0.00} & \text{\small 6.81} \text{\tiny ±0.00} & \text{\small 21.55} \text{\tiny ±0.00} & \text{\small 41.72} \text{\tiny ±0.00} \\
Linear Regression & \text{\small 3.68} \text{\tiny ±0.00} & \text{\small 4.73} \text{\tiny ±0.00} & \text{\small 3.85} \text{\tiny ±0.00} & \text{\small 4.92} \text{\tiny ±0.00} & \text{\small 21.57} \text{\tiny ±0.00} & \text{\small 30.01} \text{\tiny ±0.00} \\
AutoSARIMAX & \text{\small 7.22} \text{\tiny ±0.00} & \text{\small 9.78} \text{\tiny ±0.00} & \text{\small 4.33} \text{\tiny ±0.00} & \text{\small 5.55} \text{\tiny ±0.00} & \text{\small 56.43} \text{\tiny ±0.00} & \text{\small 76.92} \text{\tiny ±0.00} \\
Catboost & \text{\small 3.99} \text{\tiny ±0.05} & \text{\small 5.09} \text{\tiny ±0.07} & \text{\small 3.67} \text{\tiny ±0.04} & \text{\small 4.66} \text{\tiny ±0.04} & \text{\small 12.13} \text{\tiny ±1.14} & \text{\small 17.97} \text{\tiny ±2.26} \\
EUB & \text{\small 4.24} \text{\tiny ±0.00} &  \text{\small 5.46} \text{\tiny ±0.00} & \text{\small 5.10} \text{\tiny ±0.00}& \text{\small 6.74} \text{\tiny ±0.00}& - & - \\
\cmidrule(l){1-1} \cmidrule(l){2-3} \cmidrule(l){4-5} \cmidrule(l){6-7}
BiLSTM & \text{\small 3.82} \text{\tiny ±0.19} & \text{\small 4.91} \text{\tiny ±0.25} & \text{\small 3.64} \text{\tiny ±0.12} & \text{\small 4.64} \text{\tiny ±0.15} & \text{\small 21.12} \text{\tiny ±1.65} & \text{\small 32.88} \text{\tiny ±2.68} \\
MLP & \text{\small 3.64} \text{\tiny ±0.29} & \text{\small 4.61} \text{\tiny ±0.33} & \text{\small 4.82} \text{\tiny ±0.39} & \text{\small 6.17} \text{\tiny ±0.52} & \text{\small 21.45} \text{\tiny ±0.96} & \text{\small 33.34} \text{\tiny ±1.68} \\
\cmidrule(l){1-1} \cmidrule(l){2-3} \cmidrule(l){4-5} \cmidrule(l){6-7}
DLinear & \text{\small 4.42} \text{\tiny ±0.08} & \text{\small 6.33} \text{\tiny ±0.10} & \text{\small 4.28} \text{\tiny ±0.07} & \text{\small 6.10} \text{\tiny ±0.06} & \text{\small 14.26} \text{\tiny ±0.39} & \text{\small 22.55} \text{\tiny ±0.24} \\
TiDE & \text{\small 4.41} \text{\tiny ±0.03} & \text{\small 6.31} \text{\tiny ±0.03} & \text{\small 4.29} \text{\tiny ±0.02} & \text{\small 6.14} \text{\tiny ±0.02} & \text{\small 13.42} \text{\tiny ±0.47} & \text{\small 22.08} \text{\tiny ±0.35} \\
iTransformer & \text{\small 3.50} \text{\tiny ±0.07} & \text{\small 4.50} \text{\tiny ±0.07} & \text{\small 3.51} \text{\tiny ±0.19} & \text{\small 4.48} \text{\tiny ±0.24} & \text{\small 14.71} \text{\tiny ±1.88} & \text{\small 23.39} \text{\tiny ±3.17} \\
PatchTST   & \text{\small 6.47} \text{\tiny ±0.12} & \text{\small 9.08} \text{\tiny ±0.21} & \text{\small 6.34} \text{\tiny ±0.11} & \text{\small 8.91} \text{\tiny ±0.14} & \text{\small 11.51} \text{\tiny ±0.66} & \text{\small 17.68} \text{\tiny ±0.89} \\
\cmidrule(l){1-1} \cmidrule(l){2-3} \cmidrule(l){4-5} \cmidrule(l){6-7}
CE-TST (Ours) & \text{\small 3.60} \text{\tiny ±0.04} & \text{\small 4.73} \text{\tiny ±0.07} & \text{\small 3.12} \text{\tiny ±0.19} & \text{\small 3.99} \text{\tiny ±0.20} & \text{\small 16.79} \text{\tiny ±2.10} & \text{\small 25.87} \text{\tiny ±4.61} \\
CE-STF (Ours) & \textbf{\text{\small 3.19}} \text{\tiny ±0.04} & \textbf{\text{\small 4.14}} \text{\tiny ±0.07} & \textbf{\text{\small 3.09}} \text{\tiny ±0.13} & \textbf{\text{\small 3.95}} \text{\tiny ±0.17} & \text{\small 11.93} \text{\tiny ±0.96} & \text{\small 19.87} \text{\tiny ±1.96} \\
CE-CF (Ours) & \text{\small 3.80} \text{\tiny ±0.13} & \text{\small 4.89} \text{\tiny ±0.15} & \text{\small 3.22} \text{\tiny ±0.18} & \text{\small 4.11} \text{\tiny ±0.20} & \textbf{\text{\small 5.50}} \text{\tiny ±0.37} & \textbf{\text{\small 7.73}} \text{\tiny ±0.44} \\
\bottomrule
\end{tabular}
\end{table}

\section{Discussion}

The findings of this study underscore the importance of integrating  \textit{expected future} information  into load forecasting models to achieve enhanced accuracy and robustness.

By achieving significant performance improvements in both  Swiss National Railway Traction Network and building energy systems case studies, our contextually enhanced transformer framework demonstrates its effectiveness across diverse and complex energy environments. These results highlight that leveraging detailed planning and scheduling information, alongside traditional data sources like weather forecasts, can substantially reduce forecasting errors and mitigate the occurrence of large outliers. This advancement is particularly vital in the context of increasingly decentralized energy systems, where precise local forecasts are essential for the efficient operation of flexibility markets and the integration of distributed energy resources
\cite{plaumAggregatedDemandsideEnergy2022}.

Moreover, the economic implications of our improved forecasting accuracy are significant. Utility companies can realize substantial cost savings by optimizing energy procurement and reducing  reliance on expensive intra-day trading. Notably, even a 1\% reduction in forecasting error can result in annual savings of up to \$1.6 million \cite{anandBottomupForecastingApplications2023}. From an environmental perspective, more reliable forecasts facilitate the seamless integration of renewable energy sources, promoting cleaner and more sustainable energy systems while reducing the dependence on resource-intensive storage technologies \cite{ahmadReviewRenewableEnergy2020}.  Our findings demonstrate that by integrating future contextual information into transformer-based forecasting models, we can substantially enhance forecast accuracy and robustness, addressing the critical need for reliable energy predictions in increasingly decentralized and complex energy landscapes.

The findings of this study highlight the necessity of collecting and integrating diverse types of contextual information to advance load forecasting models. Our contextually enhanced transformer approach not only outperforms traditional  methods in these complex, multi-dimensional forecasting scenarios but also offers significant economic and operational benefits.

Our analysis reveals  that while state-of-the-art  multi-step models perform well on datasets characterized by clear trends and periodicity,
they are less effective in scenarios that require \textit{expected future }information  for load forecasting.
Our work demonstrates that the simultaneous integration of the full sequence of \textit{expected future }
information in a multi-step model is more beneficial than the previous paradigm of stepwise incorporation 
of exogenous variables into autoregressive models, such as SARIMAX,
but challenge the prevailing notion established in recent literature that simpler, multi-step linear forecasting methods are  universally effective
\cite{zengAreTransformersEffective2022, dasLongtermForecastingTiDE2023}. 
In our work, we evaluate  three effective embedding strategies for integrating future contextual information.
We provide empirical evidence from two distinct case studies: a novel, complex, multi-year railway load forcasting case studies introduced in this work  as our primary focus
and  a well-established multi-year building energy dataset used to validate our approach used as an auxiliary case study to validate the broader applicability of our approach.
 These comprehensive case studies highlight the versatility and robustness of our framework in different real-world applications. The railway load forecasting showcases our model's ability to handle complex, large-scale, and dynamic data, demonstrating its effectiveness in a novel and challenging domain. Meanwhile, the building energy dataset validation serves as an auxiliary case, reinforcing the framework’s applicability in a more traditional and widely studied domain.

Furthermore, our research on transformer models reveals that despite their complexity, the advanced patching and embedding strategies employed in the most recent models
such as PatchTST or iTransformer, do not consistently yield superior outcomes when contextual information impacts the performance. 
Surprisingly, for transformers, simpler linear embeddings that directly integrate future contextual information often outperform  more complex methods.
This finding  highlights the need to reevaluate embedding strategies in forecasting applications,  emphasizing the substantial   impact that directly integrating future contextual information has on enriching model inputs. Consequently, 
we demonstrate that this integration enhances forecasting accuracy and robustness across diverse and complex datasets, often surpassing the effects of advanced embedding and trend decomposition strategies in the first layers of the model.

While our study demonstrates the effectiveness of contextually enhanced transformer models, there remain avenues for further improvement. Future research can explore robust forecasting and probabilistic forecasting to enhance model resilience against uncertainties and provide probabilistic estimates of future loads. Robust forecasting techniques can improve model performance in the face of anomalous data or unexpected events, while probabilistic forecasting can offer a range of possible outcomes, aiding in better decision-making under uncertainty. Additionally, investigating hybrid models that combine the strengths of transformers with other advanced machine learning techniques could yield even greater improvements in forecasting accuracy and reliability.

By addressing  forecasting demands of decentralized energy systems and flexibility markets our research paves the way for more accurate, reliable, and economically beneficial energy management practices. The generalizability demonstrated through our two case studies and three datasets affirms the broad applicability of our framework, promising significant contributions to the field of energy forecasting.

These results advocate for the broader application of transformer-based models in various forecasting tasks due to their broad transferability. For example, beyond energy systems, similar approaches can be applied to financial forecasting by integrating market trends and economic indicators, or to supply chain management by incorporating inventory schedules and demand projections. This transferability demonstrates the versatility of our framework in addressing the nuanced demands of multilayered and complex data environments across different sectors.

\section{Methods}

We reconceptualize  multi-step time-series forecasting
as a combined forecasting and regression problem, leveraging both 
historical data and 
rich, timetable-based \textit{future contextual information}.
In this work, we propose to address this challenge through a sequence modeling approach using transformer models enriched with contextual information.
The transformer \cite{vaswaniAttentionAllYou2017}
has achieved remarkable success in natural language processing.
Since then, it has also become a foundation model for computer vision \cite{dosovitskiyImageWorth16x162021a},
and time-series \cite{dasDecoderonlyFoundationModel2024},
as it adheres to the scaling law where larger models will continue to perform better \cite{kaplanScalingLawsNeural2020}.
Building on the inherent  strengths of transformers, 
our approach modifies its architecture 
to more effectively manage and integrate  complex temporal contexts.
Our proposed approach adapts the transformer architecture 
to compute a sequence of predictions 
that integrate both past and future contextual information.
We specifically employ  encoder-decoder style transformers  for this purpose.

{\small
\textbf{Notation}: In this work, we use slicing notation denoted using the colon (\(:\)) symbol.
For a matrix \( A \in \mathbb{R}^{m \times n} \), where \( m \) and \( n \) denote  the number of rows and columns, respectively, slicing is denoted by \(A[i:j, k:l]\) or \(A_{i:j, k:l}\), where the indices \( i \) through \( j-1 \) select rows and  \( k \) through \( l-1 \) select columns of matrix \( A \). The omission of \( i \) or \( k \)  implies selection  starting from the first row or column, while  the omission of  \( j \) or \( l \) extends the selection to the last row or column.
We use $\otimes$ to denote element wise multiplication and
$\oplus$ for concatenation. The Frobenius norm is represented by
\(\Vert \bullet \Vert_{\text{F}}\).
}

\subsection{Efficient integration of the expected future in Forecasting}
\label{sec:efficient_integration_of_expected_future}

A model that effectively integrates  historical data with the \textit{expected future}
must be capable of simultaneously forecast 
based on past data and regressing from anticipated future conditions.
To date, only a few algorithms approach forecasting as a dual task, 
combining both forecasting and regression. This dual approach is essential for capturing the complex dynamics of both historical trends and future anticipated information.
Typically, forecasting models rely on historical data patterns and often lack the flexibility or capability to integrate dynamic, forward-looking inputs effectively.
While conventional regression models and
 statistical analysis methods 
can  model the relationship between covariates and the target  effectively
\cite{hongEnergyForecastingReview2020},
they often fail to adequately account for historical dependencies.
Conversely, recent state-of-the-art
multiple-input multiple-output
linear models, such as 
Decomposition Linear (DLinear, \cite{zengAreTransformersEffective2022}),
and conventional auto-regressive models like LSTMs,
do not explicitly address the regression task \cite{zhengElectricLoadForecasting2017}.
In addition, recent time-series transformer models,
such as Crossformer \cite{zhangCrossformerTransformerUtilizing2022} 
and iTransformer \cite{liuITransformerInvertedTransformers2023},
focus on broad applicability
but do not emphasize the integration of  detailed exogenous 
multivariate time series representing the \textit{expected future}
\cite{ahmedTransformersTimeSeriesAnalysis2023}. 
Although the concept of
\textit{expected future inputs} was first 
introduced in transformers by
\citet{limTemporalFusionTransformers2021}, these newer models have yet to fully exploit this approach to enhance forecasting accuracy through the integration of anticipated future conditions.

In the load forecasting literature, the integration of contextual information concerns data about socio-economic factors, weather variables, measurement about the state of the energy system and time index. In the case of short-time load forecasting, the most commonly used contextual information are weather and time index variables \cite{kusterElectricalLoadForecasting2017}. A traditional forecasting method represents the similar day method \cite{hong_probabilistic_2016}, where the most similar historical day is chosen as forecast. Thereby, the similarity measure involves the comparison of time index and weather patterns. Another traditional time series method is the Seasonal Autoregressive Integrated Moving Average with eXogenous Input (SARIMAX) \cite{weron_modeling_2007}. It takes historical load and contextual information as input, and the orders for seasonality, autoregressive lags, differencing and moving average are determined by classical time series analysis. The forecast is then produced with a recursive forecasting strategy, where the previous load forecast is fed in a loop across the forecasting horizon while also providing future contextual information such as weather forecasts. Further common approaches are regression-based techniques include Linear Regression, Support Vector Machine or Gradient Boosting Regression Trees \cite{hong_probabilistic_2016}. Deep Learning methods like Multilayer Perceptron (MLP), Long Short-Term Memory (LSTM) or Convolution Neural Network (CNN) have also been explored for load forecasting \cite{ahmad_load_2022}. However, for both regression-based and Deep Learning models, most works opt for a direct forecasting strategy by learning a mapping directly from features to future values of the target variable. This does not allow for an integration of weather forecasts directly across the forecasting horizon.
In summary, the majority of load forecasting studies do rely on weather variables and time index as contextual information. Nevertheless, apart from SARIMAX studies, weather forecasts are mainly integrated as additional input feature, neglecting its potential of direct integration across the forecasting horizon. Furthermore, the use of additional future contextual information, such as measurements about the energy system's conditions, is under-explored.

Numerous transformer architectures have been proposed 
to enhance state-of-the-art performance
in various time-series analysis tasks
\cite{wenTransformersTimeSeries2023}. Particularly,
specialized studies  have made significant advancements in energy forecasting scenarios 
using time-series transformers
\cite{ranShorttermLoadForecasting2023}.
In our work, we further develop time-series transformers
by incorporating elements from regression transformers.
Previous research on regression transformers has shown 
that a range of  regression problems  
can be approached as conditional sequence learning tasks. 
Notable examples include 
symbolic regression
\cite{kamiennyEndtoendSymbolicRegression2022},
linear regression
\cite{pathakTransformersCanOptimally2023},
and applications in 
computational chemistry 
\cite{irwinChemformerPretrainedTransformer2022}
\cite{bornRegressionTransformerEnables2023}.
Building  on the concept  that sequence modeling 
principles are applicable to time series analysis
\cite{nieTimeSeriesWorth2022},
we propose a novel and effective  approach  to
enhance encoder-decoder-based transformer models. Our strategy involves integrating covariates 
from the \textit{expected future} by 
modifying the  embedding layer of the transformer's decoder. 
In this innovative approach, the decoder functions as the regressor, 
selectively attending to sequence data from the \textit{expected future},
while the encoder learns representations of  past data.
We enhance both components  by
introducing an additional trainable embedding at each time step to capture  the \textit{expected future}. 
Departing from traditional transformer architecture,
our model  employs  non-causal attention, enabling  it
to leverage the embedded information across  all time steps for more effective forecasting. 
This dual  formulation as a regression and forecasting task not only   improves  generalization
 capabilities beyond  standard  forecasting methods  but also reduces   
overfitting in smaller datasets -- 
a notable challenge with conventional transformers.
By reframing  the forecasting problem in this dual manner, our approach reduces  the dependency  on long input sequences for context interpretation, effectively 
addressing  a common   limitation of transformer models, which struggle with  handling  long input contexts 
\cite{ahamedTimeMachineTimeSeries2024}.

\subsection{Problem Formulation}
\label{sec:problem_formulation}

Let \(X_i[\mathsf{t-w:t}] = \{x_{t-w}, \dots, x_{t-1}\}\)
represent the $i$-th input sequence 
and \(X_i[\mathsf{t:t+h}] = \{x_t, x_{t+1}, \dots, x_{t+h}\}\)
the associated target sequence which lies in the future
-- the grid load in this work --
with a past window of length $w$, a forecasting horizon $h$.
The target sequence is \(D_t\)-dimensional such that \( x_t \in \mathbb{R}^{D_t}\).
Similarly, we denote  the associated 
tabular, contextual information from the past as:
\(C_i^p[\mathsf{t-w:t}] = \{c_{t-w}^p, \dots c_{t-1}^p\} \), 
\(c_t^p \in \mathbb{R}^{D_c^p}\) 
and the \textit{future contextual information} as:
\(C_i^f[\mathsf{t:t+h}] = \{c_t^f, c_{t+1}^f, \dots, c_{t+h}^f\} \), 
\(c_t^f \in \mathbb{R}^{D_c^f}\)  
for each time step in the past and future.
Here \(D_c^p\) and \(D_c^f\) are the number of past and future covariates, respectively.
We use these definitions to introduce the \textbf{time-series regression task} where we predict the grid load \(\tilde{X_i}[\mathsf{t:t+h}]\)
by simultaneously considering the regression problem 
\(   \tilde{X_i}[\mathsf{t:t+h}] = f_r \left(C_i^f\right) \) 
and the forecasting problem
\(\tilde{X_i}[\mathsf{t:t+h}] = f_f  \left(X_i[\mathsf{t-w:t}]\right)  \).
In this work, we define a unified, parametrized
forecasting model $M_\theta$ as a function
\(\tilde{X_i}[\mathsf{t:t+h}] = M_\theta\left(X_i[\mathsf{t-w:t}], C_i^p, C_i^f\right) \) 
for a specific point in time $w < t < T$, 
predicting the grid load  \(\tilde{X_i}[\mathsf{t:t+h}]\).

\subsection{Transformer for the timeseries regression task}
\label{sec:proposed_method}

For model $M_\theta$, we propose to adapt an encoder-decoder transformer architecture 
where the encoder processes the past
and the decoder processes
the \textit{future contextual information}.
The transformer operates on a sequence of embedding vectors (embeddings).
Since \textit{future contextual information} embeddings
differ from the past grid load embeddings,
we adopt  the strategy to separate the future contextual sequence \(C_i^f\)
from the past sequence \(C_i^p\)
and train specialized 
encoder
\(f_f\) 
and decoder \(f_r\) contextual embedding layers. Specifically, we use:
\begin{equation}
\begin{array}{cc}
    \text{Encoder:} & Z_i = f_f\left(\mathtt{Embed}_c\left(C_i^p\right) \bullet \mathtt{Embed}_x\left(X_i[\mathsf{t-w:t}]\right) \right) \\
    \text{Decoder:} & \tilde{X_i}[\mathsf{t:t+h}] = f_r\left(\mathtt{Embed}_c\left( C_i^f\right),  Z_i\right)\\
\end{array}
\end{equation}
      
where \(\bullet\) represents a monoidal composition.
In this setup, the decoder serves as the regressor, 
by using \textit{non-causal attention} to attend
to data from the \textit{expected future},
while the encoder learns a representation of the past data.

\textbf{Non-Causal Attention:}
In our experiments, we adopt   non-causal (bi-directional) attention, as introduced by \citet{devlinBERTPretrainingDeep2019} in the BERT model. This choice is motivated  by the non-causal nature of our time-series regression task, 
where  the entire context -- both past and future --  is accessible at any point during the forecasting period.
By leveraging bi-directional attention, we effectively utilize all available data, enabling more comprehensive   integration of contextual information to enhance forecasting accuracy.
This approach contrasts with  causal attention mechanisms,  which prevent unavailable future tokens 
from influencing  the prediction of current tokens. Causal attention ensures that the model only uses information available at the time of prediction. However, in our context, where future contextual information is available and beneficial, non-causal attention can provide an  advantage.
This is typically achieved through a masking technique:

\begin{equation}
\text{Attention}(Q, K, V) = \text{softmax}\left(\frac{QK^T + O}{\sqrt{d_k}}\right)V
\end{equation}

Here, \(O_{ij}\)  is set to \( -\infty\) \text{ if } \(j > i\)  (for future tokens) and \(0\) otherwise,
effectively ignoring positions \(j\) that are greater than \(i\).
By modifying the mask \(O_{ij} = 0\) for all \(i\) and \(j\), 
the model
can leverage the full  bidirectional future context. This adjustment  
enhances the model's  ability to integrate information across the entire input sequence, enabling it to utilize both past and future data effectively.
The Spacetimeformer, which emerged as  the best performing model in our tests, 
leverages the permutation invariance property  of  self-attention.  
This allows it to flatten the multivariate time series, extending the attention across  all $N_i \times w$ tokens in the encoder and $N_i \times h$ tokens
in the decoder, respectively.

\subsection{Contextual Embedding}
\label{sec:st_embedding}

The inherent attention mechanism of the traditional transformer model 
is invariant to  the order of input sequences; 
however, time series data  fundamentally relies on sequentiality.
To address this, extensive research has been dedicated to developing  temporal and positional encoding strategies 
has been that reintroduce the concept of sequence into  transformers. The predominant method has been the use of 
 additive embeddings
\cite{dufterPositionInformationTransformers2022}.
Research has highlighted that the absence of positional and temporal embeddings can lead to a significant increase in forecasting
errors in transformer models \cite{zengAreTransformersEffective2022}. This suggests the critical role these embeddings play in improving the accuracy of time-series predictions made by transformers.
Despite these advancements, there remains  no universally accepted  strategy for encoding time series data, leading to varied results among different transformer models. For instance, models like the  Crossformer
have shown decreased forecasting performance when  additional covariates  are embedded 
\cite{zhangCrossformerTransformerUtilizing2022}.
This divergence highlights the ongoing debate and experimentation on the optimal way to incorporate timesemantics of time into data embeddings.
Several innovative  embedding strategies have been proposed to overcome these challenges, ranging from variable selection networks and LSTM preprocessing 
\cite{limTemporalFusionTransformers2020}
to convolutional preprocessing 
\cite{zhouInformerEfficientTransformer2021a}.
More recent approaches  have modified the self-attention mechanism  itself, employing techniques such as full dimension-wise embeddings  in the iTransformer \cite{liuITransformerInvertedTransformers2023}
and dimension-segment-wise embeddings  in the Crossformer \cite{zhangCrossformerTransformerUtilizing2022}. These methods reflect the ongoing evolution and diversity in embedding strategies, underscoring the complex nature of effectively capturing time series semantics within transformer architectures.

\textbf{Rich contextual embedding vectors:}
Given the absence of a universal embedding strategy,
we propose a method for embedding \textit{future contextual information} by replicating the value embedding technique  used in each respective  transformer model.
We modify the encoder and decoder embeddings for the   
Spacetimeformer, 
Crossformer,
and Timeseries Transformer
 as detailed in Table \ref{tab:embedding-strategy}.
Depending on the model, we apply either  summation 
or the concatenation operation \(\oplus\)
to integrate contextual information with additional embeddings 
such as positional and temporal embeddings, to enhance the model's understanding of the data. 

\begin{table}[ht]
\centering
\caption{
Comparison of  embedding strategies for the proposed contextually enhanced transformer models. \(C_i^p\) and \(C_i^f\) denote past and future contextual information respectively, while \(C_t\) represents  a periodic representation of time, and \(E_{pos}\) is the positional embedding of the token. The DSW layer is a feature  of Crossformer introduced by its creators.
}
\label{tab:embedding-strategy}
\begin{tabular}{lll}
\toprule
\textbf{Name}  
& Encoder Embedding (\(\textbf{E}_{enc}\))
& Decoder Embedding (\(\textbf{E}_{dec}\))\\ \midrule
STF & 
\(\text{Linear}(x_i \oplus \text{Linear}(C_i^p) \oplus \text{Linear}(C_t)) + E_{\text{pos}}\) &
\(\text{Linear}(\text{Linear}(C_i^f)  \oplus \text{Linear}(C_t))+ E_{\text{pos}}\)
\\ 
CF &
\(\text{LayerNorm}(\text{DSW}(X_i \oplus  C_i^p  \oplus C_t) + E_{\text{pos}})\) &
\(\text{LayerNorm}(\text{DSW}(C_i^f  \oplus C_t) + E_{\text{pos}})\)
\\ 
TST & 
\(\text{Conv1d}(X_i) + \text{Conv1d}(C_i^p)  + E_{\text{pos}}\) &
\(\text{Conv1d}(C_i^f)  + E_{\text{pos}}\)
\\ 
\bottomrule
\end{tabular}
\end{table}

We maintain  the same embedding dimensions across all models, with the exception of the Spacetimeformer. In this model, each token encapsulates  a scalar value \(x_i\)
that spans across signals and time. For the Spacetimeformer, 
we concatenate these scalar values with the value embeddings, and a final linear
layer projects them to the standard  embedding dimension.
Additionally, we enrich the models with 
temporal embeddings (\(C_t\)), such as hour-of-the-day, day-of-the-week, calendar week and month, along with 
 positional embeddings that denote the sequence position (\(E_{\text{pos}}\)). 
These enhancements are implemented in accordance with the established practices in transformer architectures.

\subsection{Training Objective}
\label{sec:training_objective}

To train our transformer model, we use a specific portion of the dataset,  designated  as
the training dataset, denoted by  
$\mathcal{D}^{\text{Train}}$. We also set up analogous comparable validation and test datasets,  represented as 
$\mathcal{D}^{\text{Val}}$ and $\mathcal{D}^{\text{Test}}$, respectively.
Although  the future contextual information is only
provided daily for the upcoming 24 hours starting at 00:00h,
we expand  the training dataset 
by implementing  a striding strategy with an hourly step size.
The model is optimized based on minimizing the forecasting error across all context windows of the forecasting horizon.  
The training objective formulated  as follows:
\begin{equation}
    \mathcal{L}_\theta(\mathbf{X}, \mathbf{C}) = \| M_\theta\left( \mathbf{X}[\mathsf{t-w:t}], \mathbf{C} \right) - \mathbf{X}[\mathsf{t:t+h}] \|_F^2  
\end{equation}

During the  training process, our goal is to determine  the optimal set of parameters \(\theta^*\) for the model \(M_\theta\) that minimizes the expected loss. This is achieved using gradient descent, as formulated below:

\begin{equation}
\theta^* = \arg\min_{\theta} \mathbb{E}_{X,C \sim \mathcal{D}_{\text{Train}}} \left[ \mathcal{L}_\theta\left(\mathbf{X},\mathbf{C} ; \theta\right) \right]
\end{equation}

We use the performance metrics derived from $\mathcal{D}^{\text{Val}}$ to guide the  adjustment and optimization of hyper-parameters.

\subsection{Datasets}
\label{sec:datasets}

In our work, we investigate  the  challenges of integrating contextual information for forecasting, using four datasets across  two distinct  environments. We highlight how variations in available  information impact  model performance, examining two datasets without contextual data  and two datasets with contextual information at varying  spatial resolution.
The \textit{Railway} datasets were collected to support grid operators and energy traders in the day-ahead energy market. 
This market is crucial for the \textit{wholesale electricity and power sector}, allowing traders to submit their bids and offers for the  electricity delivery 
for each hour of the following  day before the market closes 
\cite{weronElectricityPriceForecasting2014}.
Additionally, predictive models developed using these datasets are designed to anticipate
demand surges and facilitate efficient  load management, thereby contributing significantly to the stability and efficiency of energy markets.

\label{sec:case_study}

\textbf{Railway / Railway-agg:}
The \textit{Railway} datasets utilized in this research were  derived from this RTN and were obtained 
in collaboration with SBB specifically  for load forecasting.
The grid load within these datasets is defined as the boundary integral net input from power plants, neighboring networks and frequency converters.
From  2018 to 2023, we compiled  two comprehensive multi-year datasets that include measurements  
of the grid load along with a  rich set of covariates.
Transport-related covariates are derived from SBB's internal 
operational planning models, 
while weather data is sourced from weather stations or external climate models.
The future contextual information is provided
in daily intervals for the day ahead (the next 24 hours).
The \textit{Railway} dataset includes 52 covariates
of four geographic sectors (west, east, central, south). 
This regional data includes temperature readings, 
tonnage, kilometers traveled, 
gross tonne-kilometers, and train counts derived from the timetable 
for regional, long-distance or intercity, and cargo trains.
The \textit{Railway-Agg} dataset,  
a condensed variant on the national scale,
comprises 16 covariates of identical types 
derived from an alternative operational planning model.
For a detailed breakdown of the date ranges and further dataset specifics, please refer to
Appendix \ref{sec:additional_implementation_details}. 
To  enhance the understanding of  grid load dynamics within this network, 
we have included visualizations depicting  the grid load for three representative weeks 
in Appendix \ref{sec:additional_implementation_details}. These visualizations are designed to illustrate  typical load scenarios, providing a clear view of the fluctuations encountered. Additionally, they highlight the challenges associated with forecasting  these   highly varying dynamics, thereby underlining the complexity of the task at hand.

\textbf{Alphabuilding}: 
The building dataset is derived from a simulation using the medium-sized office building U.S. prototype \cite{li_synthetic_2021}. The building has three floors with total floor area of 4’890 square meters and is constructed in a vintage style following ASHRAE 90.1-2013 \cite{ashrae90.1-2013}. The building's HVAC system is assumed to meet standard efficiency requirements per ASHRAE 90.1-2013 \cite{ashrae90.1-2013} and comprises an air handling unit per floor. Each air handling unit is equipped with  an air-cooled direct expansion cooling coil and a gas heating coil.
Additionally, each thermal zone within a floor is served by a variable air volume unit with electric reheating coils. Specifications for lighting and miscellaneous electric loads are detailed in \cite{li_synthetic_2021}.
Occupancy schedules are generated using an agent-based stochastic simulator, producing dynamic occupancy profiles \cite{chen_agent-based_2018}. These profiles are correlated with operating schedules for lighting and miscellaneous electric loads, as well as thermostat setpoints, using  the OpenStudio extension Gem \cite{li_openstudio-occupant-variability-gem_2020}.
Weather data from  Miami's climate is utilized for the simulation. \\
The forecasting task is set-up as a day-ahead building energy forecast, with the original data resampled from 10-minute intervals to 30-minute intervals -- a commonly employed resolution for day-ahead load forecasting \cite{pelekis_comparative_2023}. The dataset spans the years 2002-2003 and is split evenly into 50\% training and 50\% test set to ensure comprehensive  coverage of seasonal effects. A context window of one week (336 steps) is used to predict the next day’s energy consumption (48 steps).

\subsection{Model Training and Evaluation Criteria}

For model training, we use \texttt{PyTorch} and its implementation of the \texttt{AdamW} optimizer.
Our training regimen incorporates a custom learning rate scheduling that includes a warm-up phase and reduces the learning rate upon reaching a  plateau.
We use min-max normalization.
All four datasets consist of hourly averages, and consequently,
we  forecast a horizon \(h\) of 24 time-steps 
for the day-ahead load forecast. 
Training and validation processes are detailed in  Appendix \ref{sec:additional_implementation_details}.
To preserve the integrity of the evaluation, the temporal ordering of the training, 
 validation, and test datasets is strictly maintained, -- ensuring that the indices for validation testing  are sequentially higher than those of  training. 
Our evaluation metrics include the standard deviation.
Due to the proprietary nature of the data provider's EUB model, 
we report only a single set of results for this model.
A notable  limitation of time series transformers
is  their limited  capability to inherently  decompose  trends 
and seasonality, especially with smaller datasets
\cite{zhouFedformerFrequencyEnhanced2022,
bentsenSpatioTemporalWindSpeed2023}.  
To address this, we manually apply  trend decomposition 
for the smaller \textit{Railway-Agg} dataset by subtracting a 96 time-step moving average.

\subsection{Baseline Models}
\label{sec:baseline_methods}

\textbf{Publicly Available Models}
In our analysis, we benchmark our contextually enhanced transformer models against  the current  state-of-the-art (SOTA) models in long-range
time-series forecasting.
Within  the family of transformer-based models,
we incorporate adaptations of the Spacetimeformer \cite{grigsbyLongRangeTransformersDynamic2022},
Crossformer \cite{zhangCrossformerTransformerUtilizing2022}, 
and evaluate the unique embedding strategies used by iTransformer \cite{liuITransformerInvertedTransformers2023},
and the patching strategy applied in PatchTST \cite{nieTimeSeriesWorth2022}.
Additionally, from the recent advancements in multi-step linear models, we include 
DLinear \cite{zengAreTransformersEffective2022}
and TiDE, a dense residual model noted for its effectiveness in long-term forecasting \cite{dasLongtermForecastingTiDE2023}.
We also assess the performance of more traditional  time-series forecasting techniques,
such as the bidirectional LSTM -- which  integrates future covariates effectively 
\cite{siami-naminiPerformanceLSTMBiLSTM2019} -- to provide a comprehensive comparison.
Additionally, we extend our comparison to include  popular gradient boosting  
framework CatBoost \cite{prokhorenkovaCatBoostUnbiasedBoosting2018}, which are well-regarded for their robustness and efficiency in various predictive modeling challenges.

\textbf{Parametric Linear Regression (EUB)}
We used the SBB proprietary  forecasting  model, known as EUB, as the baseline for our predictions.
This parametric linear regression model integrates  multiple data sources, including weather forecasts, gross-ton kilometers, temperature predictions,
as well as national and international workdays and holidays, 
coupled with historical data.
The development of the EUB model benefits from the
deep  expertise  of SBB traders in forecasting, incorporating their insights into the dynamics of load variations influenced by regional, national and international public holidays in Switzerland and neighbouring countries.  
Based on the observation that load patterns  from Tuesdays to Thursdays are generally similar, while Mondays, Fridays, Saturdays, and Sundays exhibit  distinct  characteristics \cite{boschPrognosenLeistungsbedarfsVolatiler2017}, EUB uses data from several preceding similar days to establish day-ahead forecasts. 
Due to proprietary constraints, we are unable to publish the detailed workings of the model and its complete data sources.

\textbf{Benchmark Models}
In our comparison we utilize several benchmark models ranging from naive to standard deep learning models. As naive benchmark, we take the weekly seasonal naive method due to the weekly recurrent patterns in load profiles. This method uses the load profile of the same weekday from the previous week as day-ahead forecast. In case of traditional methods, we provide the results for Linear Regression and AutoSARIMAX \footnote{The timestamp information is not included due to the explicit time series modeling. Also note that in the building energy case study, the total occupancy feature had to be removed in order to avoid rank deficiency. As a result of SARIMAX's structure, there is no result only considering past context as additional information, since exogenous variables must be included for the whole time sequence}. Furthermore, we apply K-nearest neighbors regression as representative of similarity-based methods and a Multilayer Perceptron as a standard deep learning model. For K-nearest neighbor regression, we set the number of nearest neighbors to 5. The Multilayer Perceptron has two layers of 100 neurons each and is trained with the Adam optimizer, employing early stopping with a patience of 10.
Excluding the seasonal naive method, we employ a historical window of 1 week to predict the next day for all benchmark models. When considering the future context information in the respective scenarios, the future context information is added as additional features.

\subsubsection*{Data Availability}
All of the data that support the findings of this study are available with the software bundle.

\subsubsection*{Code Availability}
We provide step-by-step instructions to use the contextually enhanced 
transformer for the time series regression task in the software bundle. 
The code will be openly available on Github on publication of the paper.

\subsubsection*{Acknowledgments}
This research was funded by the Swiss Federal Office of Transport (FOT) under the project INtellIgenT maIntenance rAilway power sysTEms (INITIATE). The authors would like to thank FOT for the project coordination and Swiss Federal Railways (SBB) for providing the data for this research and the discussions on the research results and the paper. 

\subsubsection*{Declaration of Generative AI and AI-assisted technologies in the
writing process}
During the preparation of this work the authors used ChatGPT as language
polishing service in order to improve the readability and
clarity. After using this tool/service, the authors reviewed and edited
the content as needed and take full responsibility for the content of the
publication.

\bibliography{main}
\bibliographystyle{iclr2024_conference}

\appendix

\section{Background: Building Energy}
Load forecasting for buildings is often underestimated  within  the power engineering community, 
despite   buildings accounting for approximately 40\% of total energy use \cite{hongEnergyForecastingReview2020, zhaoReviewPredictionBuilding2012a}. 
Over  the past  decades, around one third of the global energy consumption has been attributed to building operation \cite{iea_buildings}.
In the context of global efforts towards international climate targets, energy efficiency measures for buildings are an essential tool. However, building energy systems  display considerable complexity due to the various energy types and building characteristics \cite{zhao_review_2012}. In order to identify and implement appropriate measures, accurate building load forecasting is an integral part for building owners and energy system operators. On the building level, load forecasts can assist in three substantial ways. Firstly, the energy cost management can be improved by identifying future usage patterns and implement demand reduction strategies or retrofitting schemes. Secondly, the building operation efficiency can be enhanced through accurate predictions for scheduling and control of HVAC systems and consequently reduce energy waste. Thirdly, the integration with local renewable energy generation can be aligned with energy consumption by maximizing self sufficiency. The effect of these assisting approaches are especially effective in the case of larger buildings such as office buildings.
On the urban level, building load forecasts aid for long-term power grid planning, supply-demand balancing or incentivization of demand response programs \cite{kazmiTenQuestionsConcerning2023}. \\
Building load forecasts are mainly influenced by ambient weather conditions, historical demand as well as exogenous factors such as occupant behavior. While weather conditions and historical demand lay a solid forecasting foundation, information about indoor conditions like occupant behavior can be beneficial since it reflects the building usage time and intensity \cite{amayri_estimating_2019, newsham_building-level_2010, tang_establishment_2020}. Thus, integrating future occupant behavior can provide useful information to the forecasting model about expected building usage.

\section{Additional Experimental Results \& Insights }
\label{sec:additional_results_and_insights}

\textbf{Multi-step Linear Models:}
This study contributes to the ongoing discussion in the field of deep-learning based forecasting,  
specifically addressing recent debates on the effectiveness of transformers for time-series forecasting.
Recent studies have indicated  that
linear models, such as DLinear or 
Time-series Dense Encoder (TiDE, \cite{dasLongtermForecastingTiDE2023})
often outperform transformers in many scenarios.
However, while TiDE and DLinear perform well in straightforward forecasting tasks, 
our research  reveals that their  advantage diminishes in scenarios that heavily depend on rich future contextual information. 
In such contexts, the enhanced capability of our proposed transformer model to integrate and leverage detailed future covariates effectively mitigates the performance edge of these linear models.
Moreover, our transformer models maintain 
competitive performance with state-of-the-art results  
on standard time-series benchmark datasets such as
ETTh1 and ETTh2, which do not incorporate expected future contextual information.   
This demonstrates  their strong generalization capabilities and affirms their viability and effectiveness in the domain of time-series forecasting.

\begin{figure}[h]
\centering
\includegraphics[width=\linewidth]{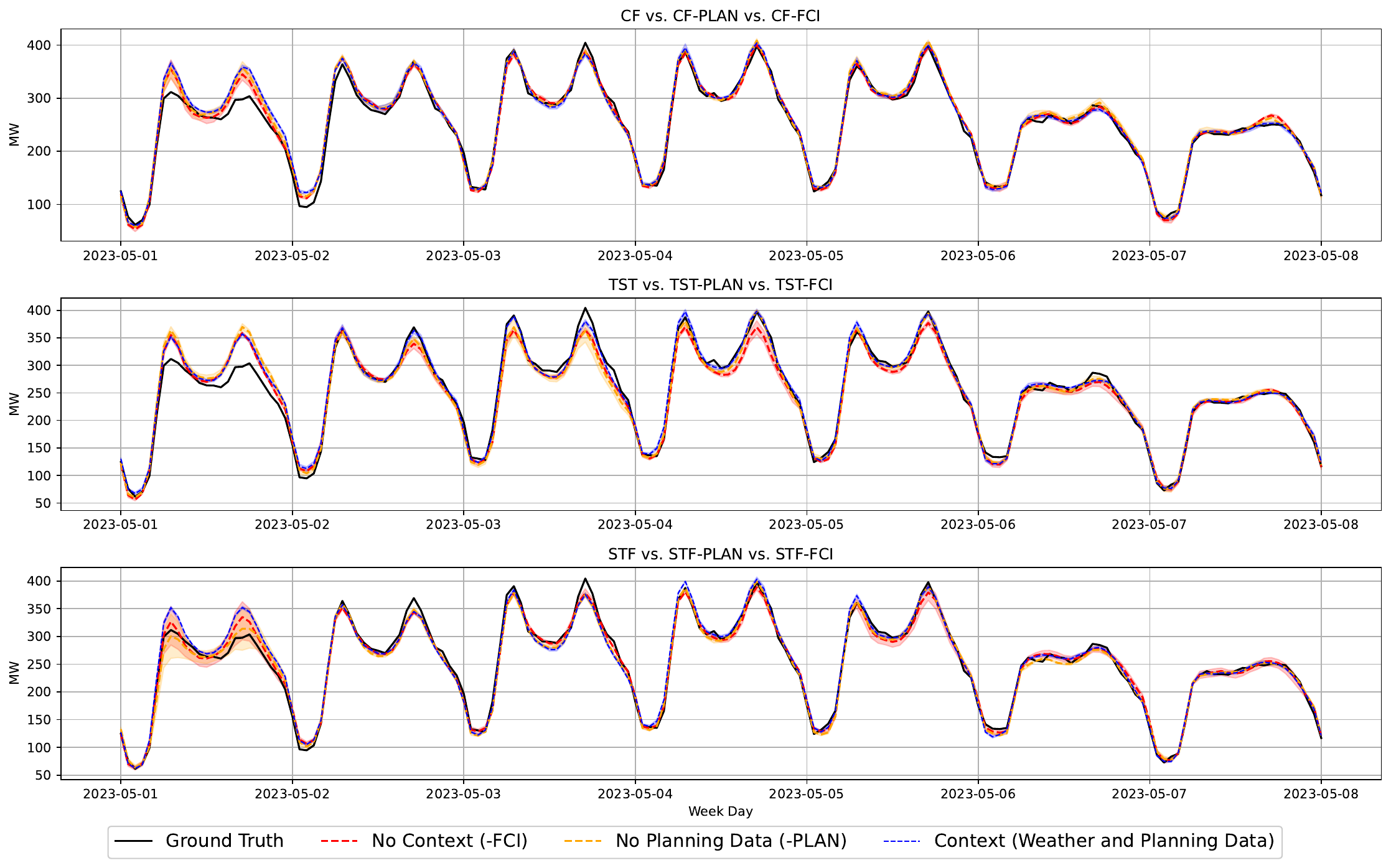}

\caption{
May 2023 (May 1st holiday)
A typical load profile overlaid with the next day forecasts (24 time steps) by contextually enhanced transformer model (CF, TST, STF).
We plot the forecast with and without \textit{future contextual information}(-FCI).
To highlight the impact of different data sources, we separately examine weather data and future planning data in the forecast plot (-PLAN).
Error bands show variation across training runs.
}
\label{fig:typical_load_profile_05_01}
\end{figure}

\begin{figure}[h]
\centering
\includegraphics[width=\linewidth]{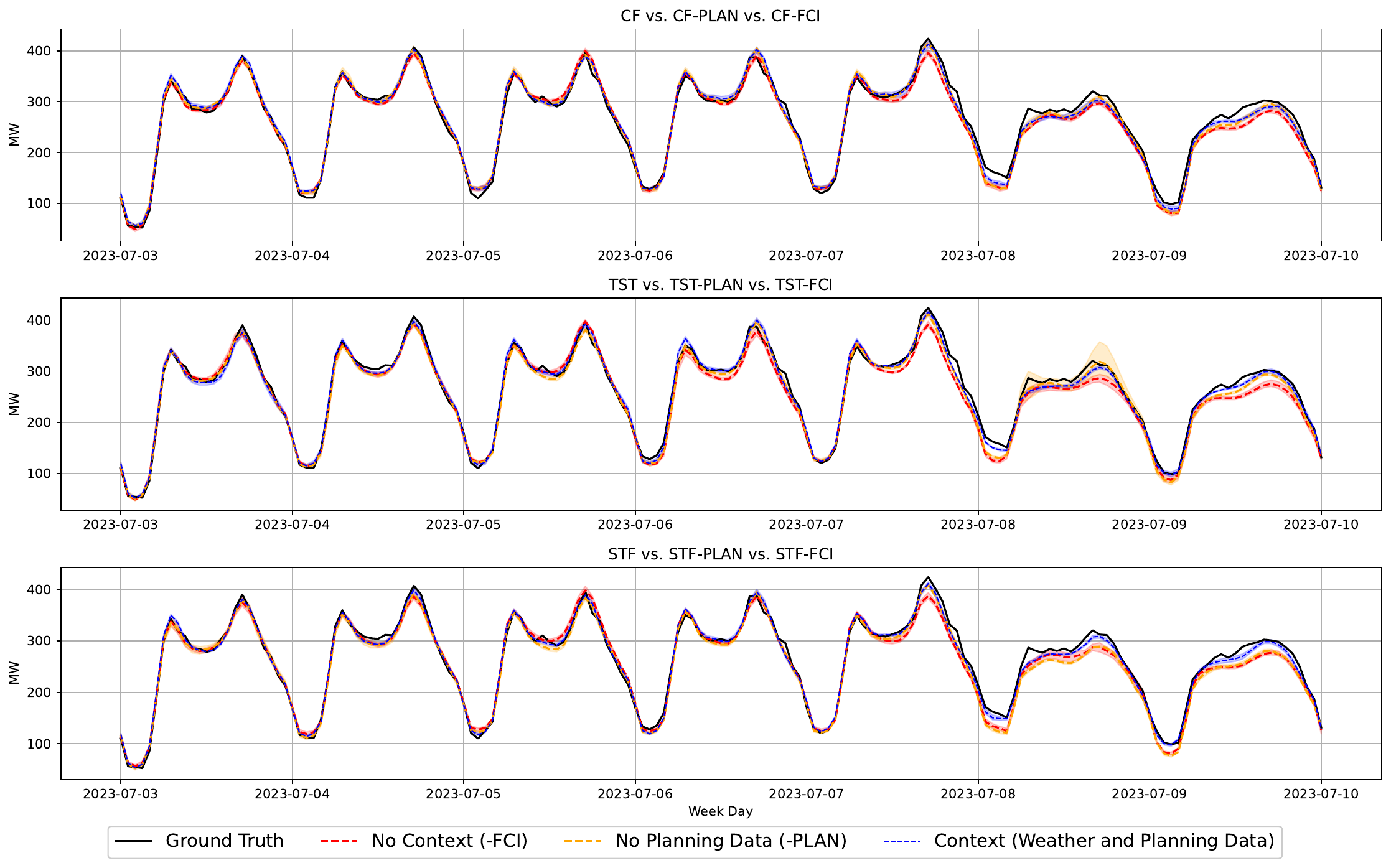}

\caption{
July 2023 (start of summer break)
A typical load profile overlaid with the next day forecasts (24 time steps) by contextually enhanced transformer model (CF, TST, STF). 
We plot the forecast with and without \textit{future contextual information}(-FCI).
To highlight the impact of different data sources, we separately examine weather data and future planning data in the forecast plot (-PLAN).
Error bands show variation across training runs.
}
\label{fig:typical_load_profile_07_03}
\end{figure}

\begin{figure}[h]
\centering
\includegraphics[width=\linewidth]{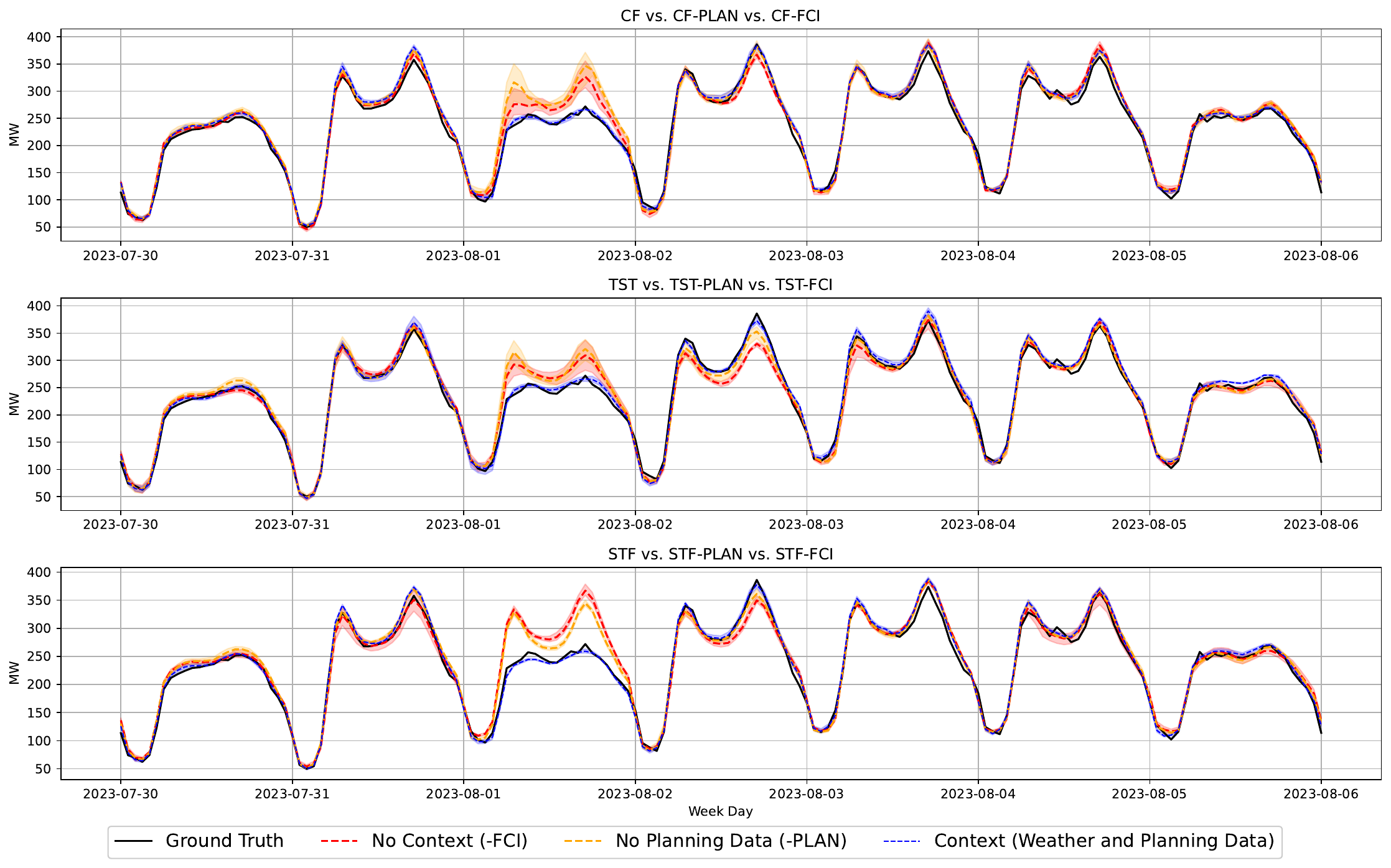}

\caption{
August 2023 (national holiday)
A typical load profile overlaid with the next day forecasts (24 time steps) by contextually enhanced transformer model (CF, TST, STF). 
We plot the forecast with and without \textit{future contextual information}(-FCI).
To highlight the impact of different data sources, we separately examine weather data and future planning data in the forecast plot (-PLAN).
Error bands show variation across training runs.
}
\label{fig:typical_load_profile_07_30}
\end{figure}

\textbf{Detailed Load Curves:}
Figure \ref{fig:typical_load_profile_05_01},
Figure \ref{fig:typical_load_profile_07_03} 
and Figure \ref{fig:typical_load_profile_07_30}
show typical load profiles for challenging contexts.
The curves emphasize the dynamic power profile 
of the Swiss traction power grid 
throughout a single day with the typical two load peaks 
during the morning and afternoon rush hour.
Curve statistics such as the load peak magnitudes 
strongly differ when comparing workdays with the weekend or with seasonal events such as holidays and vacation periods, 
requiring the model to generalize well 
to different operational contexts and seasonal conditions. 
In the detailed curves we find that the model variance is 
much larger for different random initialization
when not enhanced with \textit{future contextual information}.
Figure \ref{fig:typical_load_profile_05_01} depicts the forecast during
International Workers' Day, which is traditionally affected by strikes
and thus unpredictable. Due to this unpredictability, we expect the forecast to be 
challenging and not in line with the \textit{expected future}.
On the contrary, Figure \ref{fig:typical_load_profile_07_03} is the
regular beginning of national summer vacation time.
In this time the results support our expectation and reveal that the 
\textit{future contextual information} helps to reduce model variance
and mean prediction error.

\begin{figure}[h]
\centering
\begin{subfigure}[t]{0.48\textwidth}
\centering
\includegraphics[height=1\textwidth]{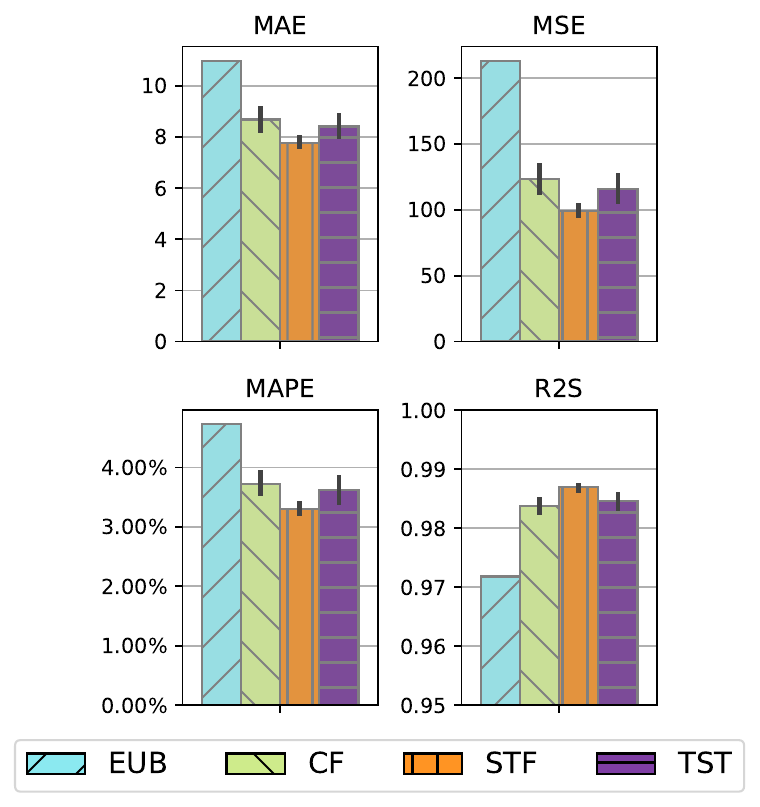}
\caption{  
Comparison of the forecasting performance.
Performance averaged over the test set. 
}
\label{fig:overall_model_performance_railway}
\end{subfigure}
\hfill
\begin{subfigure}[t]{0.48\textwidth}
\centering
\includegraphics[height=1\textwidth]{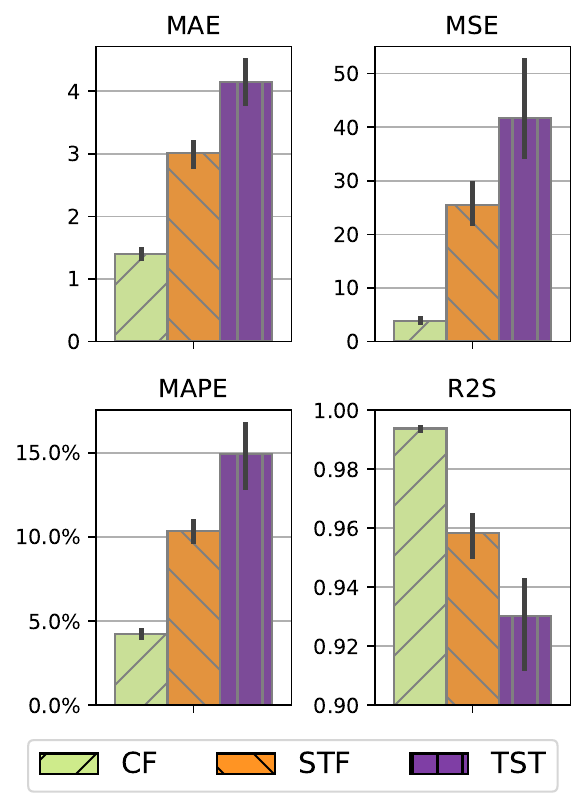}
\caption{  
Comparison of the forecasting performance.
Performance averaged over the test set. 
}
\label{fig:overall_model_performance_building_energy}
\end{subfigure}
\caption{
Comparison of the performance
of contextually enhanced transformer models: Crossformer (CF),
Spacetimeformer (STF) and Timeseries Transformer (TF)
trained and evaluated on the \textit{Railway} dataset in a) and on the \textit{Building Energy} dataset in b).
The linear regression model (EUB), 
currently the best performing model in production at the data supplier, is also included for comparison.
Error bands illustrate  the variation across different training initializations.
}
\label{fig:averaged_performance_comparison}
\end{figure}

\begin{figure}[h]
\centering
\includegraphics[width=\linewidth]{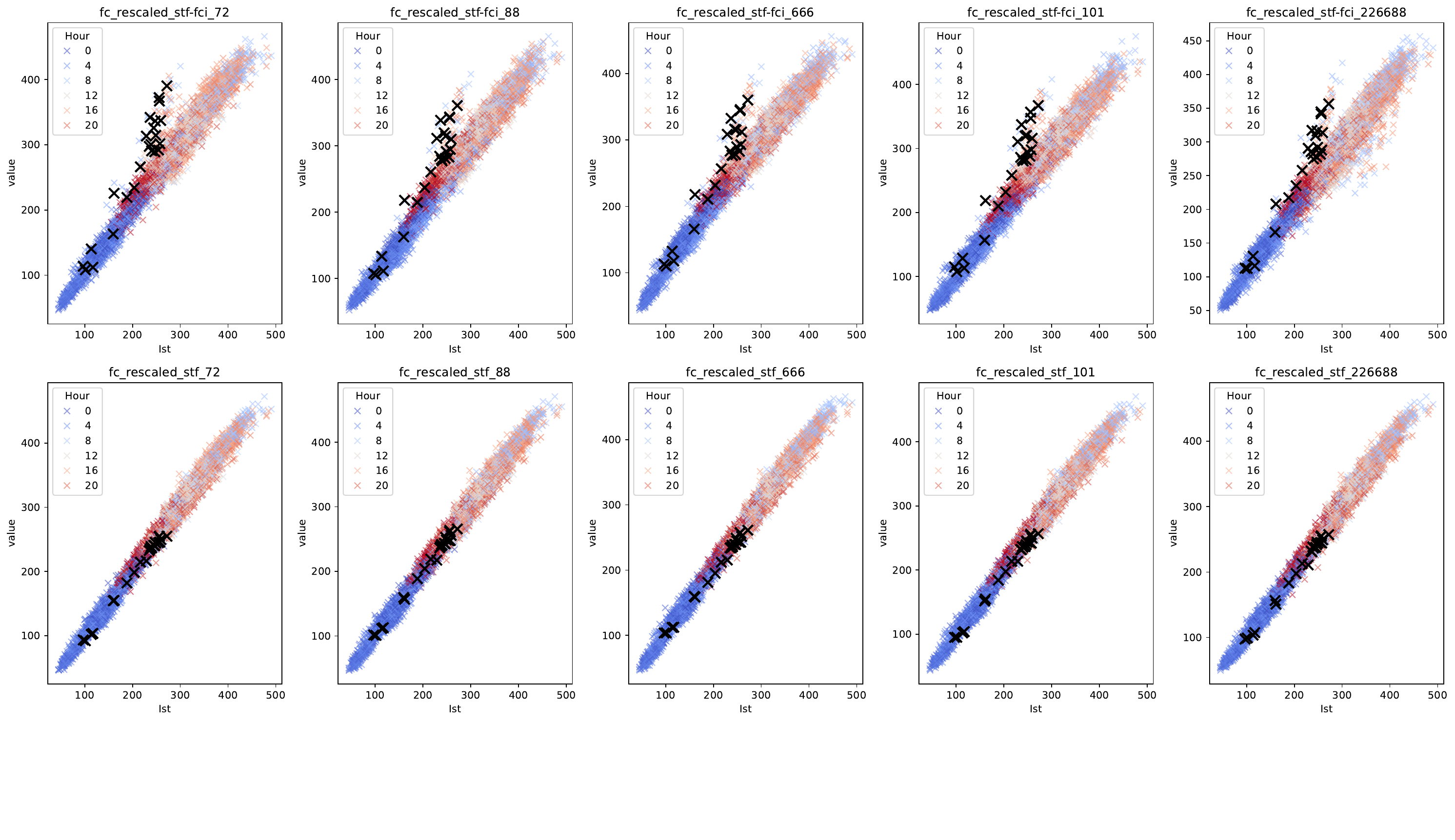}

\caption{
All additional random seeds for the Figure \textbf{Model Performance Case Study: Swiss National Holiday (August 1, 2023)}
}
\label{fig:performance_weekdays}
\end{figure}

\textbf{Performance Analysis by Weekday:} 
Further analysis  by weekday is necessary  as distinct load patterns emerge on different weekdays due to varying transportation demands and operational dynamics, which differ from weekend patterns, as detailed in Appendix
\ref{sec:additional_results_and_insights}.
Our findings indicate that while  TiDE's performance remains  unchanged with the inclusion of  FCI, all transformer models exhibit  consistent improvements across every day of the week, as depicted 
in Figure \ref{fig:performance_weekdays} . 
This  visually underscores  the beneficial impact of FCI. Notably, in the TiDE model, FCI is processed simultaneously
with past contextual information, which may dilute its effectiveness (refer to Table \ref{tab:ablations_railway} and Table \ref{tab:ablations_building} for details).
The uniform improvement across all weekdays
in transformer models highlights  the robustness and generalizability  of FCI's positive effects, demonstrating its substantial value in enhancing forecasting accuracy in complex scenarios.

\begin{figure}[h]
\centering
\includegraphics[width=\linewidth]{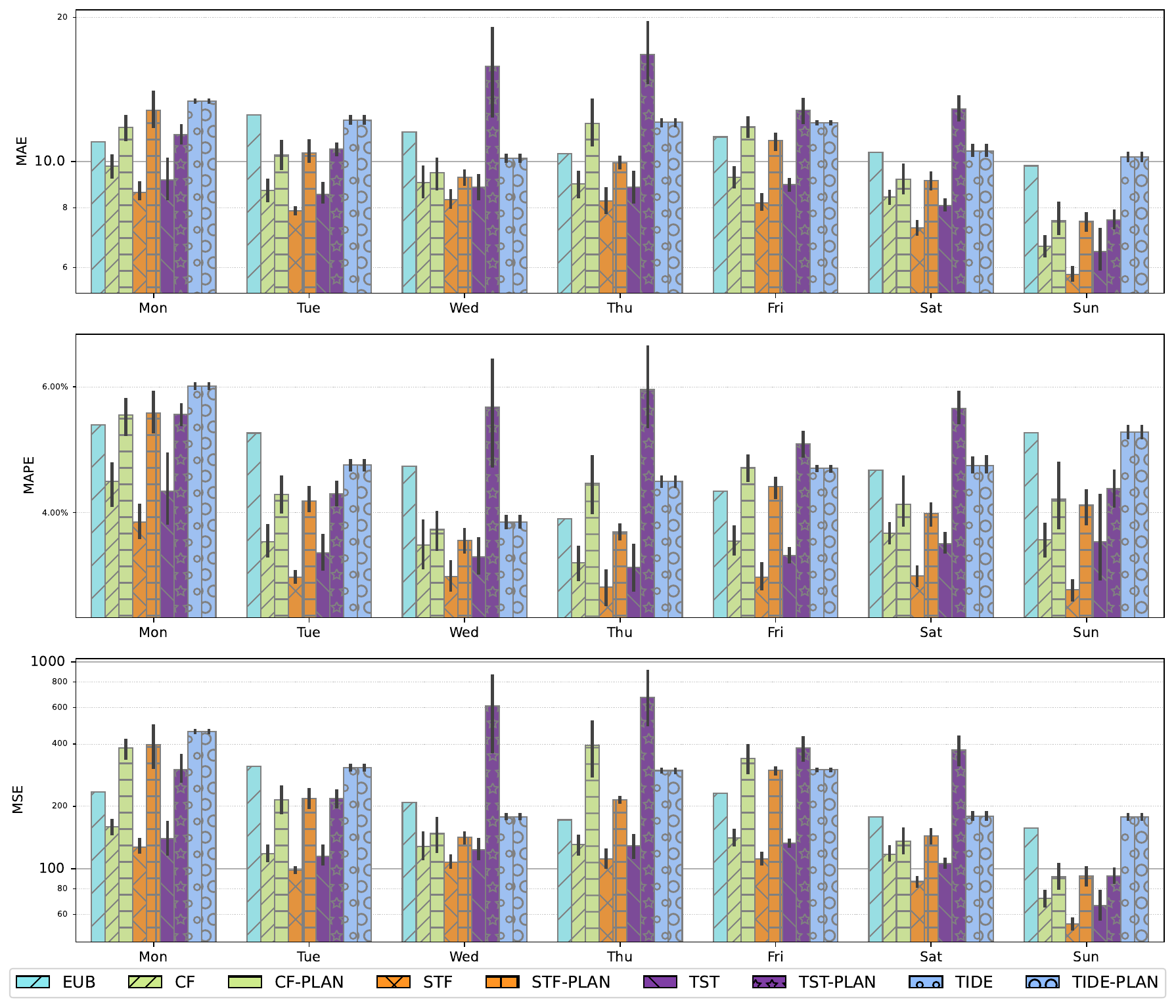}

\caption{
Comparison of the forecasting performance of contextually enhanced Crossformer (CF),
Spacetimeformer (STF), Timeseries Transformer (TST) and TiDE (TIDE)
trained on the \textit{Railway} dataset 
for each day of the week, starting on Monday on a logarithmic scale.
We also include the regression model (EUB) 
currently in production at the Swiss Federal Railways.
Performance averaged over the test set. 
We show the performance with and without (-PLAN) future planning information.
}
\label{fig:performance_weekdays_railway}
\end{figure}

\begin{figure}[h]
\centering
\includegraphics[width=\linewidth]{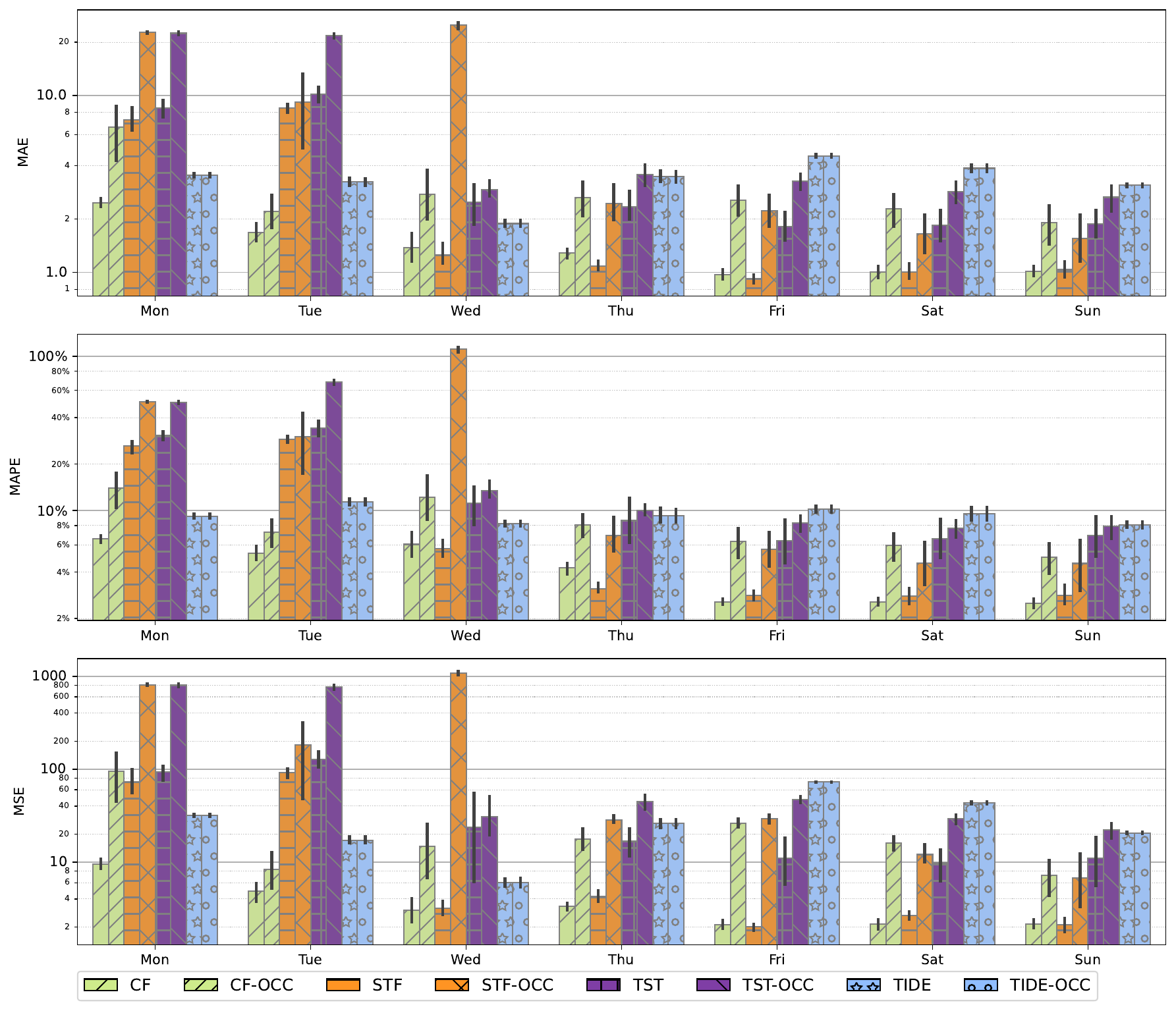}

\caption{
Comparison of the forecasting performance of contextually enhanced Crossformer (CF),
Spacetimeformer (STF), Timeseries Transformer (TST) and TiDE (TIDE)
trained on the \textit{Building Energy} dataset 
for each day of the week, starting on Monday on a logarithmic scale.
Performance averaged over the test set. 
We show the performance with and without (-OCC) future occupancy information.
}
\label{fig:performance_weekdays_alphabuilding}
\end{figure}

\begin{figure}[h]
\centering
\includegraphics[width=\linewidth]{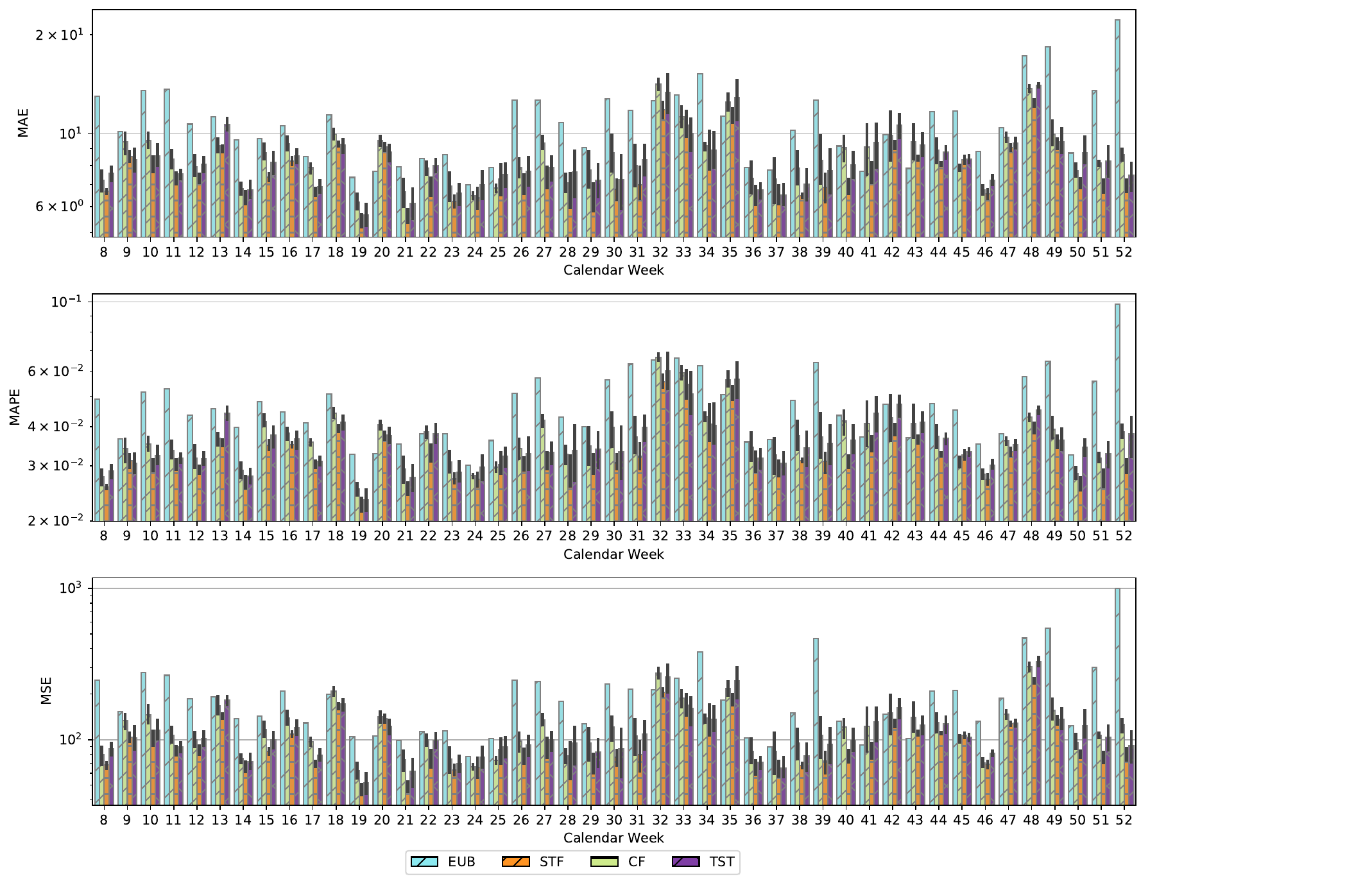}

\caption{
Comparison of the forecasting performance of contextually enhanced Crossformer (CF),
Spacetimeformer (STF), Timeseries Transformer (TST) and TiDE (TIDE)
trained on the \textit{Railway} dataset 
for each calendar week on a logarithmic scale.
We also include the regression model (EUB) 
currently in production at the Swiss Federal Railways.
We show the performance with and without (-PLAN) future planning information.
}
\label{fig:performance_per_calendar_week_railway}
\end{figure}

\begin{figure}[h]
\centering
\includegraphics[width=\linewidth]{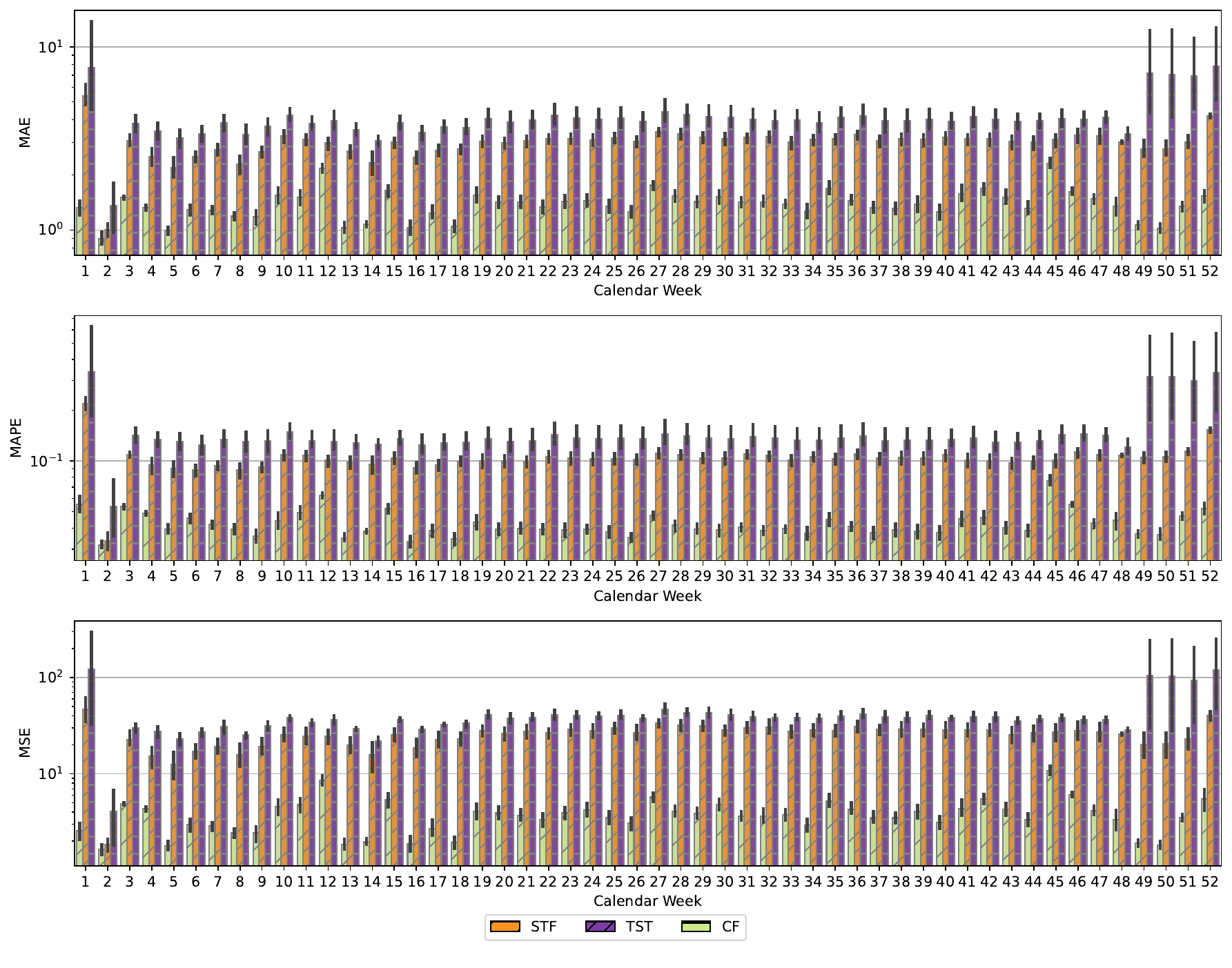}

\caption{
Comparison of the forecasting performance of contextually enhanced Crossformer (CF),
Spacetimeformer (STF), Timeseries Transformer (TST) and TiDE (TIDE)
trained on the \textit{Building Energy} dataset 
for each calendar week on a logarithmic scale.
We show the performance with and without (-PLAN) future planning information.
}
\label{fig:performance_per_calendar_week_alphabuilding}
\end{figure}

\begin{figure}[h]
\centering
\begin{subfigure}[b]{\textwidth}
\centering
\includegraphics[width=\linewidth]{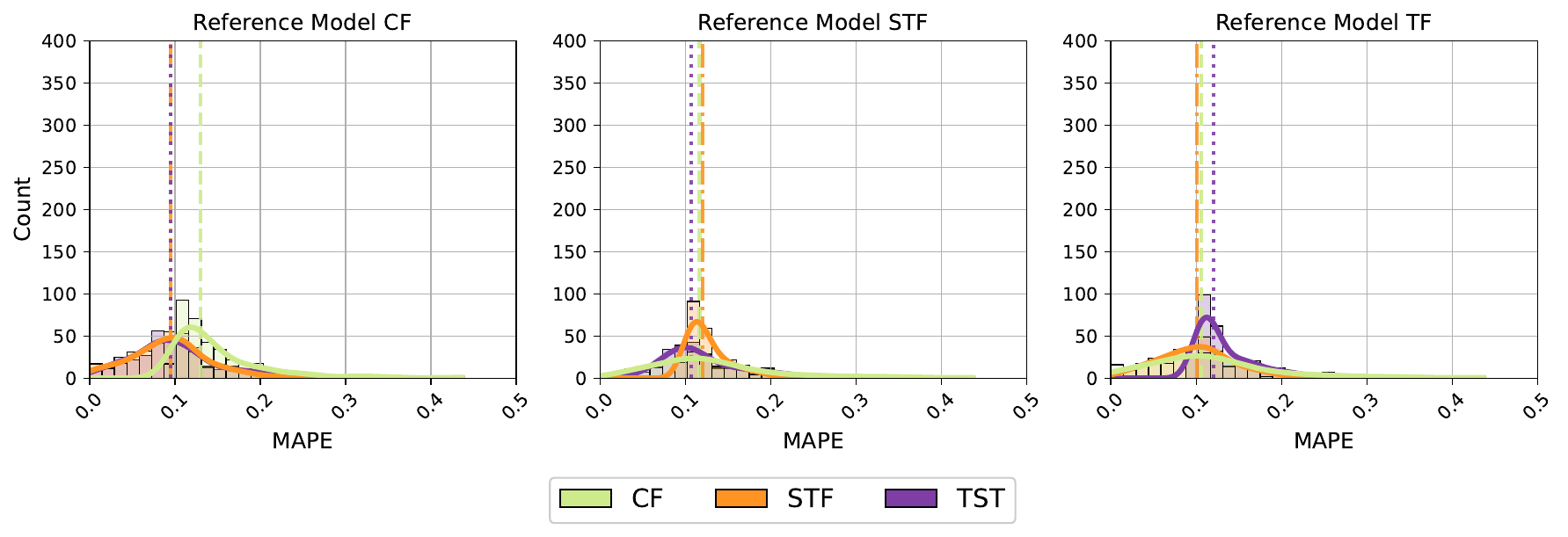}
\caption{with \textit{future contextual information}}
\vspace{1em}
\label{fig:outlier_distributions_a_railway}
\end{subfigure}
\vfill
\begin{subfigure}[b]{\textwidth}
\centering
\includegraphics[width=\linewidth]{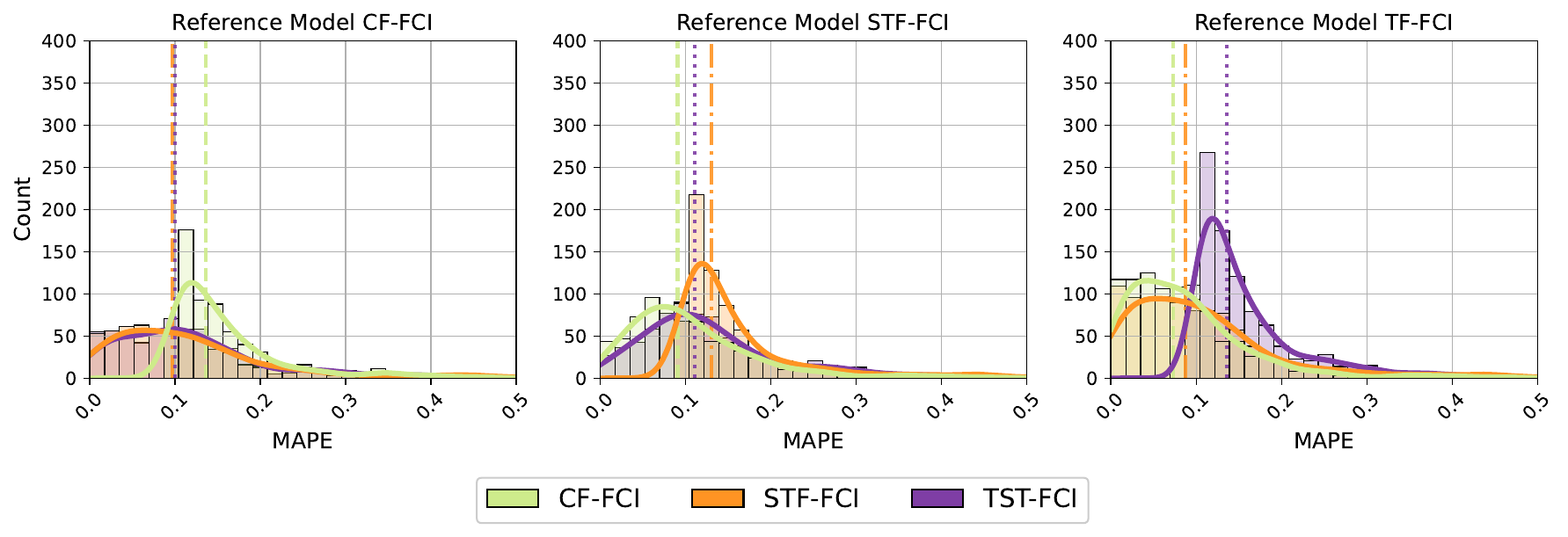}
\caption{without \textit{future contextual information}}
\label{fig:outlier_distributions_b_railway}
\end{subfigure}
\caption{
Outlier severity distributions by MAPE for the \textit{Railway} test set
for each reference model.
We select the outliers instanced where the MAPE is larger than $10\%$. We plot the median
as a vertical line.
}
\label{fig:outlier_distributions_railway}
\end{figure}

\begin{figure}[h]
\centering
\begin{subfigure}[b]{\textwidth}
\centering
\includegraphics[width=\linewidth]{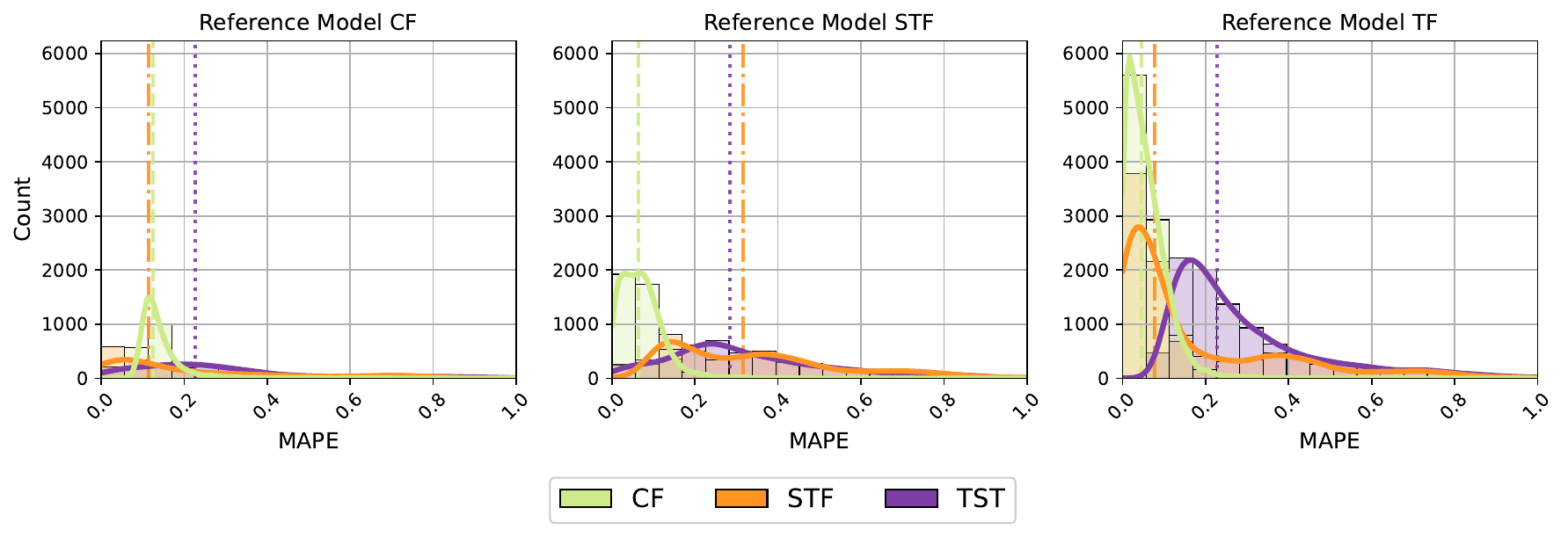}
\caption{with \textit{future contextual information}}
\vspace{1em}
\label{fig:outlier_distributions_a_alphabuilding}
\end{subfigure}
\vfill
\begin{subfigure}[b]{\textwidth}
\centering
\includegraphics[width=\linewidth]{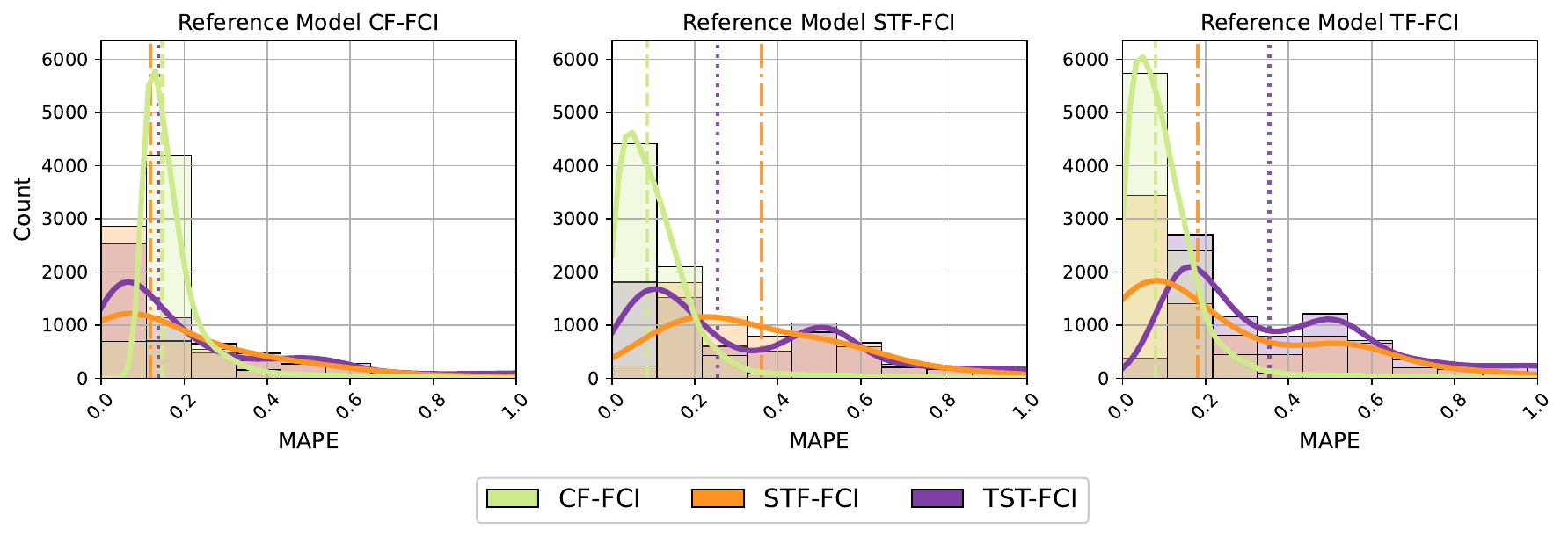}
\caption{without \textit{future contextual information}}
\label{fig:outlier_distributions_b_alphabuilding}
\end{subfigure}
\caption{
Outlier severity distributions by MAPE for the \textit{Building Energy} test set
for each reference model.
We select the outliers instanced where the MAPE is larger than $10\%$. We plot the median
as a vertical line.
}
\label{fig:outlier_distributions_alphabuilding}
\end{figure}

\textbf{Outlier Distributions:}
In an additional study we analyze 
the outlier distributions by MAPE threshold (10\%)
for each time-series transformer. 
We plot the outlier distribution for all models selected by time-stamp
of outliers of the reference model
in Figure \ref{fig:outlier_distributions_railway} and Figure  \ref{fig:outlier_distributions_alphabuilding}.
We find that without contextual
information, the other models' outlier distributions 
do not align with the reference model, 
with the distributions' means 
disagreeing with each other,
indicating the models fail to generalize in certain random situations where other models perform well.
With contextual information, the other model's distributions' mean 
is more aligned,
indicating similar (weak) 
performance on few anomalous events
inherently given by the dataset.

\begin{figure}[h]
\centering
\includegraphics[width=0.48\textwidth]{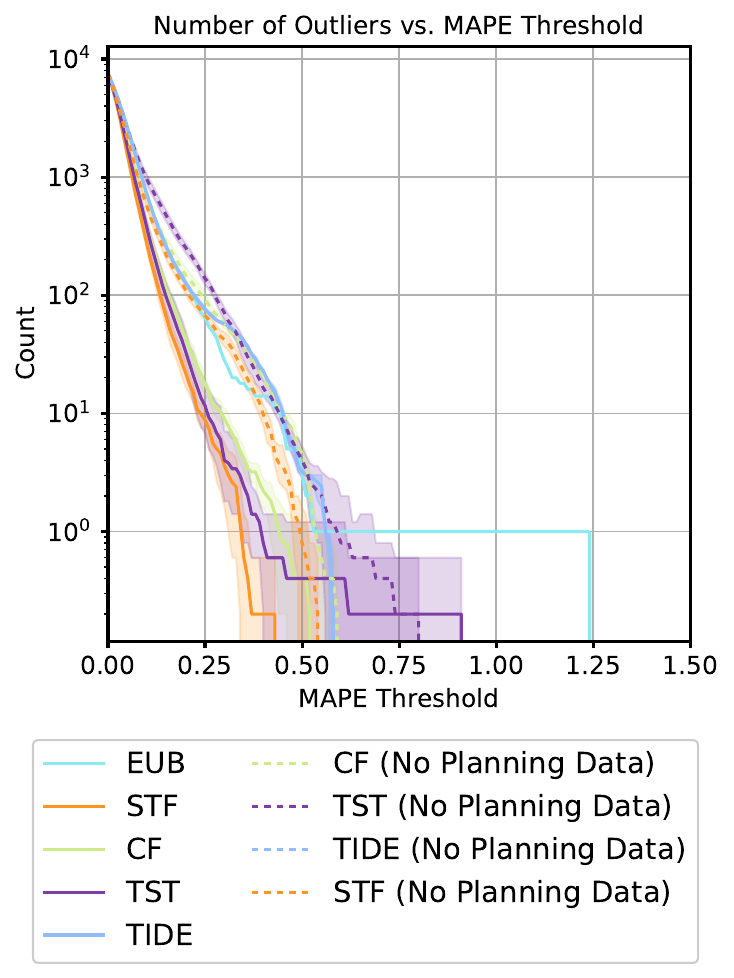}
\caption{
Outlier counts by forecasting model plotted against the MAPE threshold for the full threshold range on the \textit{Railway} test set.
}
\label{fig:mape_outliers_zoomed_out}
\end{figure}

\begin{figure}[h]
\centering
\includegraphics[width=\linewidth]{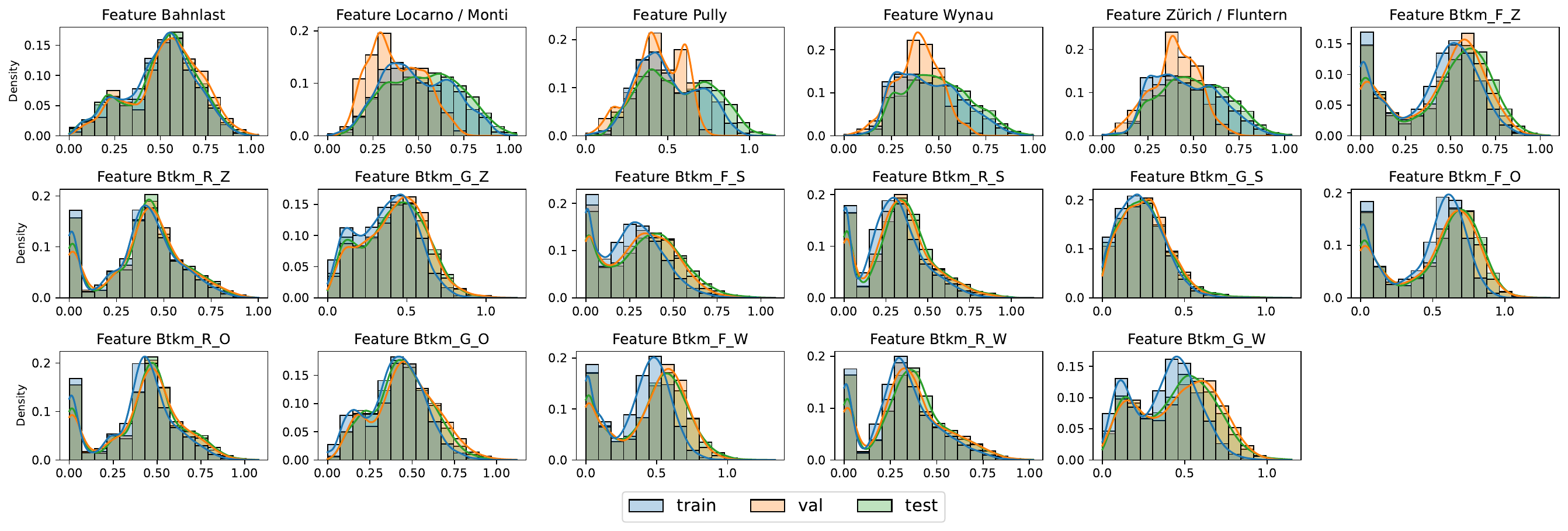}
\caption{
Data distribution shift of selected features during the COVID-19 period.
}
\label{fig:covid_distributions}
\end{figure}

\textbf{COVID-19 Case Study:}
The unforeseen onset of the pandemic lead to
a distribution shift in gross tonne-kilometres transported
on the railway network 
driven by changes in commercial and residential 
demand due to widespread lockdowns and work-from-home policies.
Therefore, the \textit{Railway} and \textit{Railway-agg} datasets present a generalization 
challenge, as they include periods during and outside of the COVID-19 pandemic.
We examine the impact of the COVID-19 pandemic on the Swiss railway traction network. 
In Figure \ref{fig:covid_distributions}, we depict the data distribution shift.
We list the forecasting performance of contextually enhanced transformers, trained exclusively on data from the pandemic period, in Table \ref{tab:covid_study} (details on data splits are listed in Table \ref{tab:data_splits}).
Both the validation set and the two test sets, Test-small and Test-large, are in the endemic phase.
We observe that the contextually enhanced transformer manages the distribution shift effectively,
experiencing a performance degradation of  
\textbf{22.1 \%} 
by mean MAE
from \textbf{8.28} to \textbf{10.11}
for out of distribution
forecasting, still surpassing the performance of the current model employed by the data supplier (EUB) which was updated weekly.

\begin{table}[h]
\centering
\caption{Results COVID-19 study}
\label{tab:covid_study}
\begin{tabular}{lllllll}
\toprule
& \multicolumn{2}{l}{Test-Small} & \multicolumn{2}{l}{Test-Large} & \multicolumn{2}{l}{Validation} \\
Ablation & MAE & MSE  & MAE & MSE  & MAE & MSE  \\
\midrule
CF & \text{\small 9.62} \text{\tiny ±0.62} & \text{\small 149.99} \text{\tiny ±18.64} & \text{\small 10.71} \text{\tiny ±0.56} & \text{\small 186.32} \text{\tiny ±16.83} & \text{\small 9.42} \text{\tiny ±0.26} & \text{\small 146.02} \text{\tiny ±6.32} \\
STF & \text{\small 9.21} \text{\tiny ±0.58} & \text{\small 135.63} \text{\tiny ±15.12} & \text{\small 10.11} \text{\tiny ±0.95} & \text{\small 166.85} \text{\tiny ±30.07} & \text{\small 8.05} \text{\tiny ±0.05} & \text{\small 105.94} \text{\tiny ±1.28} \\
TST & \text{\small 9.34} \text{\tiny ±0.67} & \text{\small 140.62} \text{\tiny ±21.56} & \text{\small 10.86} \text{\tiny ±0.96} & \text{\small 195.27} \text{\tiny ±33.40} & \text{\small 8.83} \text{\tiny ±0.25} & \text{\small 128.42} \text{\tiny ±6.43} \\
EUB & \text{\small 10.24} & \text{\small 170.42}  &\text{\small 10.95} & \text{\small 210.89} & - & - \\
\bottomrule
\end{tabular}
\end{table}

\textbf{Performance on ETTx:}
The two datasets ETT1 and ETT2 (referred to as ETTx)
without future contextual information
are commonly utilized in time series research as benchmarks for
long-range, multi-variate time-series forecasting 
\cite{zhangCrossformerTransformerUtilizing2022,
liuITransformerInvertedTransformers2023, 
zengAreTransformersEffective2022}.
The \textit{Railway} datasets, initially appearing similar to ETTx
in terms of temporal resolution (hourly) and application (power systems), differ significantly in one key aspect: while  the \textit{Railway} datasets are specifically tailored with rich contextual details, the ETTx datasets are general-purpose and relatively limited in contextual depth.

\textbf{ETTh1 / ETTh2 Dataset Information:} 
The Electricity Transformer Temperature (ETTx) datasets comprise  high-resolution time series  
capturing the operational conditions of electricity transformers,
which include oil temperature readings and six power load features \cite{zhouInformerEfficientTransformer2021a}.
For this dataset, we adopt 
the data splits and preprocessing methodologies detailed in \cite{zengAreTransformersEffective2022}.
We demonstrate  that our selected contextually enhanced transformer models achieve
competitive performance on the standard benchmark datasets ETTh1 and ETTh2. This is evidenced by their mean squared error (MSE) and mean absolute error (MAE) metrics, which 
compare favorably with SOTA timeseries forecasting models 
TiDE and DLinear, 
as detailed  in Table~\ref{tab:ettx}. 
For the Spacetimeformer, we specifically  disable the global local- and cross-attention mechanisms on the ETTh1 and ETTh2 datasets.

\textbf{ETTx Comparison of SOTA Timeseries Transformer:}
We observe that the Spacetimeformer model, 
which performs best in our study,
does not achieve the overall performance levels 
of Crossformer on ETTh1, as described in the original study, 
it however excels in the specific task of the 24-hour forecast on ETTh2, an aspect not previously listed in 
\cite{zhangCrossformerTransformerUtilizing2022}.
Interestingly,  previous  studies, such as those by Zhang et al. in 
\cite{zhangCrossformerTransformerUtilizing2022}, have noted  that segment-wise covariates could negatively impact forecasting on datasets like ETTh1.
This discrepancy might  stem from differences in the types of covariates used -- periodic covariates, 
such as hour-of-the-day and day-of-the-week, 
versus the rich contextual information used in our \textit{Railway} load forecasting.
Our approach also includes separate embeddings for periodic covariates to ensure comprehensive contextual integration.

\begin{table}[!h]
\centering
\caption{\label{tab:ettx} MAE and MSE performance on the ETTx datasets for a forecasting window of 24. We use the same training environment as for the \textit{Railway} datasets. }
\begin{tabular}{lllll}
\toprule
& \multicolumn{2}{l}{ETTh1} & \multicolumn{2}{l}{ETTh2} \\
\cmidrule(l){2-3} \cmidrule(l){4-5}
& MAE & MSE  & MAE & MSE   \\
\midrule
CF & \text{\small 0.371} \text{\tiny ±0.008} & \text{\small 0.309} \text{\tiny ±0.007} & \text{\small 0.400} \text{\tiny ±0.011} & \text{\small 0.316} \text{\tiny ±0.024} \\
DLinear & \text{\small 0.345} \text{\tiny ±0.000} & \text{\small 0.298} \text{\tiny ±0.001} & \text{\small 0.274} \text{\tiny ±0.005} & \text{\small 0.183} \text{\tiny ±0.005} \\
STF & \text{\small 0.386} \text{\tiny ±0.009} & \text{\small 0.335} \text{\tiny ±0.007} & \text{\small 0.332} \text{\tiny ±0.022} & \text{\small 0.239} \text{\tiny ±0.030} \\
TiDE & \text{\small 0.352} \text{\tiny ±0.000} & \text{\small 0.312} \text{\tiny ±0.000} & \text{\small 0.260} \text{\tiny ±0.000} & \text{\small 0.171} \text{\tiny ±0.001} \\
\bottomrule
\end{tabular}
\end{table}

\section{Additional Implementation Details}
\label{sec:additional_implementation_details}

\textbf{Evaluation Metrics:} 
We evaluate the models on MSE, MAE, MAPE and scaled versions (by $1e2$ of NMSE and NMAE):

\begin{equation}
    \text{MSE} = \frac{1}{n} \sum_{i=1}^{n} (y_i - \hat{y}_i)^2
\end{equation}
\begin{equation}
    \text{MAE} = \frac{1}{n} \sum_{i=1}^{n} |y_i - \hat{y}_i|
\end{equation}
\begin{equation}
    \text{MAPE} = \frac{1}{n} \sum_{i=1}^{n} \left| \frac{y_i - \hat{y}_i}{y_i} \right|
\end{equation}

\begin{equation}
    \text{NMSE} = 1e2 * \frac{\sum_{i=1}^N (y_i - \hat{y}_i)^2}{\sum_{i=1}^N |y_i|}
\end{equation}
    
\begin{equation}
    \text{NMAE} = 1e2 * \frac{\sum_{i=1}^N |y_i - \hat{y}_i|}{\sum_{i=1}^N |y_i|}
\end{equation}

Hyper-parameter optimization was performed using grid search,
with initial parameter ranges selected from the original works.
Hyper-parameter tuning was performed on the validation set.
We list the best hyper-parameters in Table \ref{tab:model_details}.
and the data splits in Table \ref{tab:data_splits}.

\begin{table}[!h]
\centering
\caption{
\label{tab:ablations_building} Ablation study on \textit{Building Energy}.
We list the performance in megawatts on the \textit{Building Energy} dataset. We show the input window length \(w\) and if the model has access to \textit{future contextual information} (FCI). Further, we indicate the baselines models use no \textit{contextual information} (CI) with (w/o CI).
}
\begin{tabular}{llllll}
\toprule
& \multicolumn{3}{l}{Future Context Information} & \multicolumn{2}{l}{Test Error}\\
\cmidrule(l){2-4} \cmidrule(l){5-6} 
Ablation & \(w\) & Weather + Time & Occupancy & NMAE & NRMSE   \\

\midrule
Weekly Naive (w/o CI) & 336 &\xmark &\xmark& \text{\small 10.76} \text{\tiny ±0.00} & \text{\small 21.79} \text{\tiny ±0.00} \\
KNN Regression (w/o FCI)& 336 &\xmark &\xmark & \text{\small 24.85} \text{\tiny ±0.00} & \text{\small 44.08} \text{\tiny ±0.00} \\
KNN Regression& 336 &\cmark &\cmark& \text{\small 21.55} \text{\tiny ±0.00} & \text{\small 41.72} \text{\tiny ±0.00} \\
Linear Regression (w/o FCI)& 336 &\xmark &\xmark & \text{\small 38.23} \text{\tiny ±0.00} & \text{\small 60.73} \text{\tiny ±0.00} \\
Linear Regression& 336 &\cmark &\cmark& \text{\small 21.57} \text{\tiny ±0.00} & \text{\small 30.01} \text{\tiny ±0.00} \\
AutoSARIMAX (w/o CI) & 336 &\xmark &\xmark& \text{\small 69.07} \text{\tiny ±0.00} & \text{\small 82.88} \text{\tiny ±0.00} \\
AutoSARIMAX & 336 &\cmark &\cmark& \text{\small 56.43} \text{\tiny ±0.00} & \text{\small 76.92} \text{\tiny ±0.00} \\
Catboost (w/o FCI)& 336 &\xmark &\xmark & \text{\small 19.74} \text{\tiny ±0.10} & \text{\small 31.36} \text{\tiny ±2.16} \\
Catboost& 336 &\cmark &\cmark & \text{\small 12.13} \text{\tiny ±1.14} & \text{\small 17.98} \text{\tiny ±2.26} \\
\midrule
BILSTM  & 48 &\cmark &\cmark  & \text{\small 21.12} \text{\tiny ±1.65} & \text{\small 32.88} \text{\tiny ±2.68} \\
MLP (w/o FCI) & 336 &\xmark &\xmark& \text{\small 38.01} \text{\tiny ±1.57} & \text{\small 65.89} \text{\tiny ±3.30} \\
MLP & 336 &\cmark &\cmark& \text{\small 21.45} \text{\tiny ±0.96} & \text{\small 33.34} \text{\tiny ±1.68} \\
\midrule
DLinear (w/o FCI)& 336 &\xmark &\xmark & \text{\small 14.75} \text{\tiny ±0.83} & \text{\small 23.23} \text{\tiny ±0.70} \\
DLinear (w/o OCC) & 336 &\cmark &\xmark& \text{\small 14.26} \text{\tiny ±0.39} & \text{\small 22.55} \text{\tiny ±0.24} \\
DLinear & 336 &\cmark &\cmark & \text{\small 14.26} \text{\tiny ±0.39} & \text{\small 22.55} \text{\tiny ±0.24} \\
TiDE (w/o FCI) & 336 &\xmark &\xmark& \text{\small 13.42} \text{\tiny ±0.47} & \text{\small 22.08} \text{\tiny ±0.35} \\
TiDE (w/o OCC) & 336 &\cmark &\xmark& \text{\small 13.42} \text{\tiny ±0.47} & \text{\small 22.08} \text{\tiny ±0.35} \\
TiDE & 336 &\cmark &\cmark& \text{\small 13.42} \text{\tiny ±0.47} & \text{\small 22.08} \text{\tiny ±0.35} \\
\midrule
PatchTST (w/o FCI)& 48 &\xmark &\xmark & \text{\small 11.54} \text{\tiny ±0.79} & \text{\small 17.56} \text{\tiny ±0.65} \\
PatchTST (w/o OCC)& 48 &\cmark &\xmark & \text{\small 11.31} \text{\tiny ±0.71} & \text{\small 17.60} \text{\tiny ±0.72} \\
PatchTST & 48 &\cmark &\cmark& \text{\small 11.51} \text{\tiny ±0.66} & \text{\small 17.68} \text{\tiny ±0.89} \\
iTransformer & 48 &\cmark &\cmark& \text{\small 14.70} \text{\tiny ±1.88} & \text{\small 23.39} \text{\tiny ±3.17} \\
\midrule
STF (w/o FCI) & 48 &\xmark &\xmark& \text{\small 37.57} \text{\tiny ±3.30} & \text{\small 68.89} \text{\tiny ±2.48} \\
STF (w/o OCC) & 48 &\cmark &\xmark& \text{\small 36.43} \text{\tiny ±2.49} & \text{\small 68.96} \text{\tiny ±3.37} \\
STF & 48 &\cmark &\cmark& \text{\small 11.93} \text{\tiny ±0.96} & \text{\small 19.87} \text{\tiny ±1.96} \\
TST (w/o FCI) & 48 &\xmark &\xmark& \text{\small 35.80} \text{\tiny ±1.09} & \text{\small 62.89} \text{\tiny ±0.95} \\
TST (w/o OCC) & 48 &\cmark &\xmark& \text{\small 34.03} \text{\tiny ±0.10} & \text{\small 62.21} \text{\tiny ±1.35}\\
TST &  48 &\cmark &\cmark & \text{\small 16.79} \text{\tiny ±2.10} & \text{\small 25.87} \text{\tiny ±4.61} \\
CF (w/o FCI)& 48 &\xmark &\xmark  & \text{\small 12.59} \text{\tiny ±0.86} & \text{\small 20.28} \text{\tiny ±0.95} \\
CF (w/o OCC)& 48 &\cmark &\xmark & \text{\small 11.96} \text{\tiny ±2.10} & \text{\small 20.53} \text{\tiny ±3.76} \\
CF & 48 &\cmark &\cmark & \text{\small 5.50} \text{\tiny ±0.37} & \text{\small 7.73} \text{\tiny ±0.44} \\
\bottomrule
\end{tabular}
\end{table}

\begin{table}[!h]
\centering
\caption{
\label{tab:ablations_railway} Ablation study on \textit{Railway}.
We list the performance in megawatts on the \textit{Railway} dataset. We show the input window length \(w\) and if the model has access to \textit{future contextual information} (FCI).
}
\begin{tabular}{llllll}
\toprule
& \multicolumn{3}{l}{Future Context Information} & \multicolumn{2}{l}{Test Error}\\
\cmidrule(l){2-4} \cmidrule(l){5-6} 
Ablation & \(w\) & Weather + Time & Planning   & NMAE & NRMSE   \\

\midrule
Weekly Naive (w/o CI)& 168 &\xmark &\xmark & \text{\small 6.67} \text{\tiny ±0.00} & \text{\small 9.49} \text{\tiny ±0.00} \\
KNN Regression (w/o FCI) & 168 &\xmark &\xmark& \text{\small 5.34} \text{\tiny ±0.00} & \text{\small 7.43} \text{\tiny ±0.00} \\
KNN Regression & 168 &\cmark &\cmark& \text{\small 5.05} \text{\tiny ±0.00} & \text{\small 6.81} \text{\tiny ±0.00} \\
Linear Regression (w/o FCI)& 168 &\xmark &\xmark & \text{\small 4.36} \text{\tiny ±0.00} & \text{\small 6.08} \text{\tiny ±0.00} \\
Linear Regression & 168 &\cmark &\cmark& \text{\small 3.85} \text{\tiny ±0.00} & \text{\small 4.92} \text{\tiny ±0.00} \\
AutoSARIMAX (w/o CI) & 168 &\xmark &\xmark& \text{\small 23.59} \text{\tiny ±0.00} & \text{\small 28.73} \text{\tiny ±0.00} \\
AutoSARIMAX & 168 &\cmark &\cmark& \text{\small 4.33} \text{\tiny ±0.00} & \text{\small 5.55} \text{\tiny ±0.00} \\
CatBoost& 168 &\cmark &\cmark & \text{\small 3.67} \text{\tiny ±0.04} & \text{\small 4.66} \text{\tiny ±0.04} \\
\midrule
MLP (w/o FCI) & 168 &\xmark &\xmark& \text{\small 5.22} \text{\tiny ±0.18} & \text{\small 7.01} \text{\tiny ±0.20} \\
MLP & 168 &\cmark &\cmark& \text{\small 4.82} \text{\tiny ±0.39} & \text{\small 6.17} \text{\tiny ±0.52} \\
\midrule
DLinear (24 w/o FCI) & 24 &\xmark &\xmark& \text{\small 9.28} \text{\tiny ±0.06} & \text{\small 12.89} \text{\tiny ±0.07} \\
DLinear (w/o FCI)& 672 &\xmark &\xmark & \text{\small 4.27} \text{\tiny ±0.06} & \text{\small 6.10} \text{\tiny ±0.06} \\
DLinear (w/o PLAN)& 672 &\cmark &\xmark & \text{\small 4.28} \text{\tiny ±0.07} & \text{\small 6.10} \text{\tiny ±0.06} \\
DLinear 24 & 24 &\cmark &\cmark& \text{\small 9.28} \text{\tiny ±0.04} & \text{\small 12.92} \text{\tiny ±0.02} \\
DLinear& 672 &\cmark &\cmark & \text{\small 4.28} \text{\tiny ±0.07} & \text{\small 6.10} \text{\tiny ±0.06} \\
TiDE (24 w/o FCI)& 24 &\xmark &\xmark & \text{\small 9.62} \text{\tiny ±0.04} & \text{\small 14.23} \text{\tiny ±0.02} \\
TiDE (w/o FCI) & 672 &\xmark &\xmark& \text{\small 4.29} \text{\tiny ±0.02} & \text{\small 6.14} \text{\tiny ±0.02} \\
TiDE (w/o PLAN)& 672 &\cmark &\xmark & \text{\small 4.29} \text{\tiny ±0.02} & \text{\small 6.14} \text{\tiny ±0.02} \\
TiDE 24& 24 &\cmark &\cmark & \text{\small 9.63} \text{\tiny ±0.05} & \text{\small 14.23} \text{\tiny ±0.02} \\
TiDE & 672 &\cmark &\cmark & \text{\small 4.29} \text{\tiny ±0.02} & \text{\small 6.14} \text{\tiny ±0.02} \\
\midrule
PatchTST (w/o FCI) &48 &\xmark &\xmark & \text{\small 5.67} \text{\tiny ±0.85} & \text{\small 8.05} \text{\tiny ±1.22} \\
PatchTST (w/o PLAN) &48 &\cmark &\xmark & \text{\small 6.20} \text{\tiny ±0.16} & \text{\small 8.71} \text{\tiny ±0.12} \\
PatchTST & 48 &\cmark &\cmark &\text{\small 6.34} \text{\tiny ±0.11} & \text{\small 8.91} \text{\tiny ±0.14} \\
\midrule
CF (w/o FCI) & 24 &\xmark &\xmark& \text{\small 6.43} \text{\tiny ±0.13} & \text{\small 9.38} \text{\tiny ±0.07} \\
CF (w/o PLAN)& 24 &\cmark &\xmark& \text{\small 3.79} \text{\tiny ±0.16} & \text{\small 5.70} \text{\tiny ±0.31} \\
CF & 24 &\cmark &\cmark& \text{\small 3.22} \text{\tiny ±0.18} & \text{\small 4.11} \text{\tiny ±0.20} \\
TST (w/o FCI)& 24 &\xmark &\xmark & \text{\small 5.23} \text{\tiny ±0.47} & \text{\small 7.53} \text{\tiny ±0.81} \\
TST (w/o PLAN)& 24 &\cmark &\xmark & \text{\small 4.63} \text{\tiny ±0.38} & \text{\small 7.13} \text{\tiny ±0.78} \\
TST & 24 &\cmark &\cmark& \text{\small 3.12} \text{\tiny ±0.19} & \text{\small 3.99} \text{\tiny ±0.20} \\
STF & 192 &\cmark &\cmark & \text{\small 4.11} \text{\tiny ±0.27} & \text{\small 5.23} \text{\tiny ±0.36} \\
STF (w/o FCI)& 24 &\xmark &\xmark & \text{\small 4.21} \text{\tiny ±0.24} & \text{\small 5.79} \text{\tiny ±0.39} \\
STF (w/o Enc Load)& 24 &\cmark &\cmark & \text{\small 4.36} \text{\tiny ±0.35} & \text{\small 5.51} \text{\tiny ±0.42} \\
STF (w/o Enc FCI) & 24 &\cmark (\xmark~Enc) &\cmark (\xmark~Enc
)& \text{\small 3.19} \text{\tiny ±0.18} & \text{\small 4.06} \text{\tiny ±0.22} \\
STF (w/o PLAN) & 24 &\cmark &\xmark& \text{\small 3.70} \text{\tiny ±0.09} & \text{\small 5.39} \text{\tiny ±0.21} \\
STF (All Encoder) & 24 &\cmark &\cmark& \text{\small 3.34} \text{\tiny ±0.15} & \text{\small 4.26} \text{\tiny ±0.20} \\
STF & 24 &\cmark &\cmark& \text{\small 3.09} \text{\tiny ±0.13} & \text{\small 3.95} \text{\tiny ±0.17} \\
\bottomrule
\end{tabular}
\end{table}

\begin{table}[h]
\centering
\caption{A comparison of the different hyper-parameters after tuning
for the contextually enhanced transformer models and TiDE on the \textit{Railway} dataset and 
if changes were made for the \textit{Building Energy} dataset.
\label{tab:model_details}
}
\begin{tabular}{lcccccc}
\toprule
 & \textbf{STF} & \textbf{CF}  & \textbf{TST} & \textbf{PatchTST}  & \textbf{TiDE} & \textbf{DLinear} \\
\midrule
Model dimension & 128  & 128 & 128 & 256 & 256 (1024) & - \\
Feed-forward & 128  & 128 & 256 & 128 & 512 & -  \\
Encoder Layer & 2  & 3 & 2 & 3 & 3  & - \\
Decoder Layer & 3  & 3 & 2& - & 3 & -\\
Attention heads & 4  &  4 & 2& 4 & - & -\\
Input Window $w$ & 24  & 24 & 24& 48 & 672(336) & 672(336) \\
Forecasting Horizon $h$ & 24(48)  & 24(48) & 24(48) & 24(48) & 24(48) & 24(48)\\
\# Parameters Railway-Agg & 403k & 2373k & 50.2k&  10.5M & 12.3M& 853k \\
\# Parameters Railway & 3.3M  & 2.4M & 723k & 10.5M & 23.7M& 2.8M \\
\# Parameters Building Energy & 3.1M  & 2.6M & 559k & 15.8M & 34.4M & 831k\\
\bottomrule
\end{tabular}
\end{table}

\begin{table}[ht]
\centering
\caption{
Data splits for the \textit{Railway} and \textit{Railway-Agg} datasets including date ranges
}
\label{tab:data_splits}
\begin{tabular}{@{}llcc@{}}
\toprule
Dataset       & Split Type    & Samples & Date Range       \\ 
\midrule
Railway    & Train        & 39264 (78\%)  & 2018-04-04 - 2022-09-25   \\
                             & Validation   & 2880 (7\%)   & 2022-09-26 - 2023-01-23    \\
                             & Test-Large         & 8208 (15\%)   & 2023-01-24 - 2023-12-31    \\
                             & Test-Small   & 4512   & 2023-01-24 - 2023-07-30    \\
\addlinespace
Railway-COVID    & Train        & 14592   & 2019-12-01 - 2021-07-30   \\
                             & Validation   & 2880    & 2022-09-26 - 2023-01-23    \\
                             & Test-Large         & 8208  & 2023-01-24 - 2023-12-31    \\
                             & Test-Small   & 4512   & 2023-01-24 - 2023-07-30    \\
                             
\addlinespace
Railway-Agg & Train        & 13584 (70\%)  & 2020-12-31 - 2022-07-19	   \\
                             & Validation   & 4512 (15\%) & 2022-07-20 - 2023-01-23  \\
                             & Test-Small-Agg   & 4512 (15\%)  & 2023-01-24 - 2023-07-30    \\

\addlinespace
Building Energy & Train        & 14592 (70\%)  & 2002-01-01 - 2002-10-31   \\
                             & Validation   & 2928 (15\%) &2002-11-01  - 2002-12-31  \\
                             & Test   & 17520 (15\%)  &  2003-01-01 - 2003-12-31    \\
\bottomrule
\end{tabular}
\end{table}

\end{document}